\documentclass[twocolumn,times,tighten]{aastex61}

\usepackage{amsmath}
\usepackage{color}
\usepackage{needspace}
\usepackage{enumerate}
\usepackage{hyperref} 
\usepackage{gensymb}

\newcommand{\angstrom}{\textup{\AA}}

\newcommand{\Halpha}{{H$\alpha$}}

\newcommand{\CaII}{{\ion{Ca}{2}}}
\newcommand{\SiIV}{{\ion{Si}{4}}}

\newcommand{\OIV}{{\ion{O}{4}}}

\newcommand{\IRIS}{\textit{IRIS}}

\accepted{in ApJ. March 23, 2018}

\shortauthors{N\'obrega-Siverio et al.}

\begin{document}

%
%
\title{On the importance of the nonequilibrium ionization of \SiIV\ and \OIV\ and 
the line-of-sight in solar surges}

%
%
\correspondingauthor{D. N\'obrega-Siverio}
\email{dnobrega@iac.es, fmi@iac.es, juanms@lmsal.com}

\author[0000-0002-7788-6482]{D. N\'obrega-Siverio}
\affil{Instituto de Astrofisica de Canarias, Via Lactea, s/n, E-38205 La Laguna (Tenerife), Spain}
\affil{Department of Astrophysics, Universidad de La Laguna, E-38200 La Laguna (Tenerife), Spain}   

\author{F. Moreno-Insertis}
\affil{Instituto de Astrofisica de Canarias, Via Lactea, s/n, E-38205 La Laguna (Tenerife), Spain}
\affil{Department of Astrophysics, Universidad de La Laguna, E-38200 La Laguna (Tenerife), Spain}  

\author[0000-0002-0333-5717]{J. Mart\'inez-Sykora}  
\affil{Lockheed Martin Solar and Astrophysics Laboratory, Palo Alto, CA 94304, USA}
\affil{Bay Area Environmental Research Institute, Institute, Moffett Field, CA 94035, USA}

%
%

\begin{abstract}
Surges are ubiquitous cool ejections in the solar atmosphere that often
appear associated with transient phenomena like UV bursts or coronal
jets. Recent observations from the \textit{Interface Region Imaging
  Spectrograph} (\IRIS) show that surges, although traditionally related to
chromospheric lines, can exhibit enhanced emission in \SiIV\ with brighter spectral
profiles than for the average transition region (TR). In this
paper, we explain why surges are natural sites to show enhanced emissivity in
TR lines. We performed 2.5D radiative-MHD numerical
experiments using the Bifrost code including the nonequilibrium ionization
of silicon and oxygen. A surge is obtained as a
by-product of magnetic flux emergence; the TR enveloping the
emerged domain is strongly affected by nonequilibrium effects:
assuming statistical equilibrium would produce an absence of \SiIV\ and
\OIV\ ions in most of the region. Studying the properties of the surge plasma 
emitting in the \SiIV~1402.77~$\angstrom$ and \OIV~1401.16~$\angstrom$ lines,
we find that a) the timescales for the optically-thin losses and heat conduction
are very short, leading to departures from statistical equilibrium, and b) the surge emits in \SiIV\ more  
and has an emissivity ratio of \SiIV\ to \OIV\ larger than a standard TR.
Using synthetic spectra, we conclude the importance of line-of-sight effects: given the involved
geometry of the surge, the line-of-sight can cut the emitting layer at small angles
and/or cross it multiple times, causing prominent, spatially intermittent
brightenings both in \SiIV\ and \OIV. 
\end{abstract}


%
%
\section{Introduction}\label{sec:introduction}
The solar atmosphere contains a wide variety of chromospheric ejections that cover a large range
of scales: from the smallest ones with maximum size of a few megameters, such as penumbral 
microjets \citep[e.g.,][]{Katsukawa:2007wk, Drews:2017}, or spicules \citep[][among others]
{Hansteen+DePontieu2006, de-Pontieu:2007kl, Pereira:2012dz}; up to ejections that can reach, 
in extreme cases, several tens of megameters, like surges \citep[e.g.,][]{canfield1996, Kurokawa2007, 
Guglielmino:2010lr,YangH:2014} and macrospicules \citep{Bohlin1975, Georgakilas, Murawski2011, 
Kayshap2013}. Surges, in particular, are often associated with magnetic flux emergence from the solar 
interior. They are typically observed as darkenings in images taken in the \Halpha\ blue/red wings with line-of-sight 
(LOS) velocities of a few to several tens of km s$^{-1}$, and are usually related to other explosive 
phenomena like EUV and X-ray jets, UV bursts and Ellerman bombs \citep[see]
[hereafter NS2017, and references therein]{Nobrega-Siverio:2017a}. Although 
observationally known for several decades now, the understanding of surges has progressed 
slowly and various aspects like, e.g., their impact on
  the transition region (TR) and corona concerning the mass and energy budget, are still poorly known.

From the theoretical point of view, the first explanation of the surge phenomenon came through 2.5D 
numerical models \citep{Shibata1992a,Yokoyama:1995uq,Yokoyama:1996kx}, where a cold ejection was 
identified next to a hot jet as a consequence of a magnetic reconnection process between the magnetic 
field in plasma emerged from the interior and the preexisting coronal field. \cite{Nishizuka:2008zl} used 
a similar numerical setup to associate the surge with jet-like features seen in \CaII\ H+K observations by 
means of morphological image comparisons. Further 2.5D models that include
the formation of a cool chromospheric ejection are those of \cite{jiang2012}
(canopy-type coronal magnetic field), and \cite{YangL:2013, YangL:2018}, who
study the cool jets resulting from the interaction between moving magnetic features at
the base of their experiment and the preexisting ambient field in the
atmosphere. Turning to three dimensional models, in the magnetic flux emergence experiment of
\cite{Moreno-Insertis:2013aa}, a dense wall-like surge appeared surrounding the emerged 
region with temperatures from $10^4$~K to a few times $10^5$~K and speeds around $50$ km s$^{-1}$. 
\cite{MacTaggart2015} found similar velocities for the surges in their 3D model of flux emergence in 
small-scale active regions. The availability of a radiation-MHD code like Bifrost 
\citep{Gudiksen:2011qy} has opened up the
possibility of much more detailed modeling of the cool ejections than
before. Bifrost has a realistic treatment of the material properties of the plasma,
calculates the radiative transfer in the photosphere and chromosphere 
and includes the radiative and heat conduction entropy sources in the corona. Using that code, \cite{Nobrega-Siverio:2016}, hereafter NS2016,
argued that entropy sources play an important  
role during the surge formation and showed that a relevant fraction of the surge could not be obtained in 
previous and more idealized experiments.

The realistic treatment of surges may require an even larger degree
  of complication. The solar atmosphere is a highly dynamical environment;
the evolution sometimes occurs on short timescales that bring different
atomic species out of equilibrium ionization, thus complicating both the
modeling and the observational diagnostics
\citep[e.g.,][]{Griem:1964,Raymond:1978,Joselyn:1979, Hansteen:1993}.  For
hydrogen, for instance, using 2D numerical experiments,
\cite{Leenaarts:2007sf, Leenaarts:2011qy} found that the temperature
variations in the chromosphere can be much larger than for statistical
equilibrium (SE), which has an impact on, e.g., its coolest regions (the
so-called cool pockets). For helium, \citet{golding2014,golding:2016}
described how nonequilibrium (NEQ) ionization leads to higher temperatures
in wavefronts and lower temperatures in the gas between shocks. For heavy
elements, \cite{Bradshaw:2003,Bradshaw:2006, Bradshaw:2011, Reep:2017}
showed, through 1D hydrodynamic simulations, that there are large departures
from SE balance in cooling coronal loops, nanoflares and other impulsive
heating events that affects the EUV emissivity. Through 3D
  experiments, \cite{Olluri:2013fu} found that deduced electron densities for \OIV\ can be up to
an order of magnitude higher when NEQ effects are taken into
account.  Also in 3D, \cite{olluri:2015} discussed the importance of the NEQ ionization of 
coronal and TR lines to reproduce 
absolute intensities, line widths, ratios, among others, 
observed by, e.g., \cite{Chae:1998, Doschek:2006, Doschek:2008qy}. 
\cite{De-Pontieu2015} were able to explain the correlation between
non-thermal line broadening and intensity of TR lines only
when including NEQ ionization in their 2.5D numerical
experiments. \cite{Martinez-Sykora:2016obs} studied the statistical
properties of the ionization of silicon and oxygen in different solar
contexts: quiet Sun, coronal hole, plage, quiescent active region, and
flaring active region, finding similarities with the observed intensity
ratios only if NEQ effects are taken into account. Given their highly
time-dependent nature and the relevance of the heating and cooling mechanisms
in their evolution, surges are likely to be affected by NEQ ionization. 
Motivated by this fact, NS2017 included the NEQ
ionization of silicon to compare synthetic \SiIV\ spectra of two 2.5D
numerical experiments with surge observations obtained by 
the \textit{Interface Region Imaging Spectrograph}
\citep[\IRIS,][]{De-Pontieu:2014vn} and the Swedish 1-m Solar Telescope
\citep[SST,][]{Scharmer:2003ve}. The results showed that the experiments were
able to reproduce major features of the observed surge; nonetheless, the
theoretical aspects to understand the enhanced \SiIV\ emissivity within the
numerical surge and its properties were not addressed in that publication.

\begin{figure*}
\epsscale{1.18}
\plotone{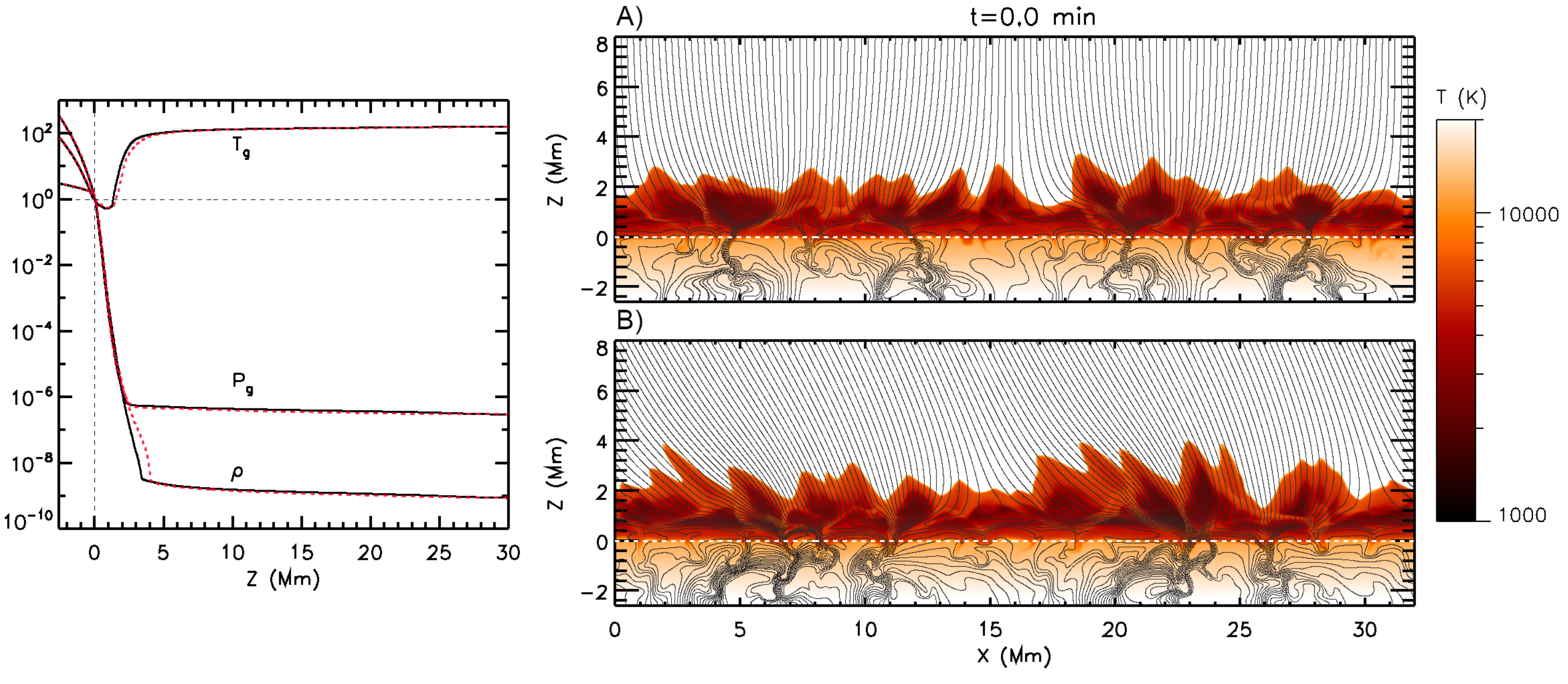}
\caption{
Left: Horizontal averages for the initial stratification of $\rho$, $P_{g}$, and $T$ normalized to their 
photospheric values at $z = 0$~Mm, namely, $\rho_{ph} = 3.1\times10^{-7}$~g cm$^{-3}$,
$P_{g_{ph}} = 1.1 \times 10^{5}$~erg cm$^{-3}$ and $T_{ph}= 5.7\times10^3$~ K. Solid black line represents the stratification for the vertical coronal field experiment; the red dotted, for the slanted one. The horizontal and vertical dotted lines mark the reference normalization values at $z=0$ Mm. Right: 2D maps for the initial temperature with magnetic field lines in black for the vertical experiment (top) and slanted experiment (bottom). The maps only show temperatures below $2\times10^4$ K (although the range varies from 1660 K up to $\sim$ 1 MK) and heights between $-2.6 \leq z \leq 8.0$~Mm (the top of the domain reaches $z=30$ Mm). The solar surface is roughly at $z=0$ Mm (white dashed horizontal line). \label{figure1}}
\end{figure*}  

The aim of the present paper is to provide theoretical explanations
concerning the relevance of the NEQ ionization for surges and the
corresponding impact on the emissivity of TR lines. We use
2.5D numerical experiments carried out with the Bifrost code
\citep{Gudiksen:2011qy} including the module developed by \cite{olluri:2013aa}
that solves the time-dependent rate equations to calculate the ionization
states of different elements, thus allowing for departures from SE. Here we apply this module to determine the
ionization levels of silicon and oxygen. We conclude that consideration of 
NEQ is necessary to get the proper population levels of the ions and, consequently,
the right emissivity to interpret observations. A statistical
analysis of temperature is provided to constrain the plasma properties
involved in the emissivity of relevant lines of \SiIV\ and \OIV\ within the
surges.  Through detailed Lagrange tracing, we are able to determine the
origin of the emitting plasma and the role of the optically thin radiation
and thermal conduction to explain the departure of SE of the relevant
ions. Furthermore, we compute synthetic profiles to understand previous
observational results and predict future ones,
highlighting the surge regions that are more likely to be detected
and addressing the importance of the angle of LOS.

The layout of the paper is as follows. Section \ref{sec:2} describes the physical and numerical models. 
Section \ref{sec:3} explains the general features of the time evolution of the experiments. In Section 
\ref{sec:4}, we show the main results of the paper splitting the section in a) the relevance of the NEQ 
ionization of \SiIV\,, and also \OIV, in surges (Section \ref{sec:4.1}); b) the consequences of the NEQ 
ionization for the surge plasma emitting in those TR lines, analyzing its properties and 
compare them with a generic quiet TR (Section \ref{sec:4.2}); and c) the origin of the NEQ plasma, 
addressing the role of the entropy sources (Section \ref{sec:4.3}). In Section
\ref{sec:5}, we have calculated absolute intensities and synthetic spectral 
profiles for diagnostic purposes and comparison with observations, emphasizing also the importance of 
the surge geometry and LOS. Finally, Section \ref{sec:6} contains a summary and conclusions.

%
%
\Needspace{5\baselineskip}
\section{The physical and numerical model}\label{sec:2}
We have run two 2.5D numerical flux emergence experiments in which surges are a natural 
consequence of magnetic reconnection processes. Those two experiments 
were also used by NS2017 and \cite{Rouppe2017} to compare the synthetic profiles with 
the complex profiles observed with \IRIS\ and SST.

This section is divided into two parts: (1) the numerical code, and (2) the description of the 
model underlying our experiments.

%
%
\Needspace{5\baselineskip}
\subsection{The numerical code}\label{sec:2.1}
The two experiments have been carried out with the 3D radiation-MHD (R-MHD)
Bifrost code \citep{Gudiksen:2011qy, Carlsson:2012uq, Hayek:2010ac}, which treats 
the radiative transfer from the photosphere to the corona and thermal conduction 
in a self-consistent manner (see also NS2016 for further details of this
code applied to surge experiments). Furthermore, we have enabled in the code a module 
developed by \citep{olluri:2013aa} to follow the NEQ ionization 
states of elements with atomic 
number greater than 2. This module solves the rate equations for those elements using 
the temperature, mass density, electronic number density $n_e$ and velocity values of the simulation without modifying the results of the R-MHD calculation, so there is no feedback, e.g., on the energy equation terms such as the optically thin losses (see the discussion in Section \ref{sec:6.1}). In particular, we have employed it to calculate the NEQ ionization fraction for silicon and oxygen, using abundances from \cite{Asplund:2009}, 7.52 and 8.69, respectively, in the customary astronomical scale
where 12 corresponds to hydrogen. 

%
%
\Needspace{5\baselineskip}
\subsection{Description of the models}\label{sec:2.2}

\Needspace{5\baselineskip}
\subsubsection{Physical domain and initial condition}

In the two experiments, we began with a statistically stationary 2D snapshot that spans
 from the uppermost layers of the solar interior to the corona, and whose physical 
 domain is $0.0$~Mm $\leq x \leq$ $32.0$~Mm and $-2.6$~Mm $\leq z \leq$ 
 $30.0$~Mm, where $z=0$~Mm corresponds to the solar surface. The grid is 
 uniform in the $x$-direction with $\Delta x=31$~km, but it is nonuniform in the 
 vertical direction in order to better resolve the photosphere and chromosphere:
 the vertical grid spacing is $20$ km  from the photosphere to the transition
 region, and increases gradually in the corona up to $147$ km at the top of the domain.

The left panel in Figure~\ref{figure1} contains the horizontal averages for the initial 
density, $\rho$, gas pressure, $P_g$, and temperature, $T$, for both experiments
normalized to photospheric values, namely, $\rho_{ph} = 3.1\times10^{-7}$~g cm$^{-3}$,
$P_{g_{ph}} = 1.1 \times 10^{5}$~erg cm$^{-3}$ and $T_{ph}= 5.7\times10^3$~ K. 
The corona has a temperature around $1$ MK and a magnetic field with a strength 
of $10$ G, with the difference that one of the experiments 
(hereafter \textit{the vertical experiment}) has a vertical 
magnetic field in the corona 
while in the other (\textit{the slanted experiment}), the magnetic field in the corona is inclined 
$30$\degree\ with respect to the vertical direction (see magnetic field lines superimposed in black 
in the 2D temperature maps for the initial snapshot in Figure~\ref{figure1}).

\Needspace{5\baselineskip}
\subsubsection{Chemical elements calculated in NEQ and their spectral lines}

We have used the NEQ module of
  \cite{olluri:2013aa}  mentioned in the \nameref{sec:introduction}  
to compute the nonequilibrium ionization of silicon 
in both numerical experiments. Furthermore, in the vertical experiment we also 
calculate the NEQ ionization of oxygen, with the goal of predicting 
future observational results.  Once the NEQ populations are obtained, we are able 
to compute the emissivity using
\begin{eqnarray}
	\epsilon_{\lambda} & = &   \frac{h\, c}{4\, \pi\, \lambda}\, n_u\, A_{ul},
 \label{eq:emissivity}
\end{eqnarray}
where $h$ is the Planck's constant, $c$ the light speed, $\lambda$ is the
wavelength of the spectral line, $n_u$ the population density of the upper
level of the transition (i.e., the number density of emitters), and $A_{ul}$ the Einstein coefficient for
spontaneous de-excitation given by
\begin{eqnarray}
	A_{ul} & = &   \frac{8\, \pi^2\, e^2\, }{m_e\, c} \frac{1}{\lambda^2}\, \frac{g_l}{g_u}\, f_{lu},
\label{eq:einstein}
\end{eqnarray}
where $e$ is the electron charge, $m_e$ the electron mass, $g_l$ and $g_u$ the statistical 
weights of the lower and upper states respectively, and $f_{lu}$ the oscillator strength. 
The units used in this paper for the emissivity $\epsilon$ are erg 
cm$^{-3}$ sr$^{-1}$ s$^{-1}$. For the sake of compactness, we will refer to it in the following 
as $\epsilon_{_{CGS}}$.

Since we are interested in understanding the response of the TR to 
chromospheric phenomena like surges, we have chosen the following \IRIS\ lines: 
\SiIV\ 1402.77 $\angstrom$, which is the weakest of the two silicon resonance 
lines; and \OIV\ 1401.16 $\angstrom$, the strongest of the forbidden oxygen lines 
that \IRIS\ is able to observe. The corresponding formation temperature peaks in 
statistical equilibrium, $T_{_{SE}}$, and other relevant parameters to calculate
the Einstein coefficient (Equation \ref{eq:einstein}) and the corresponding emissivity 
(Equation \ref{eq:emissivity}) of these lines are shown in Table \ref{table1}.
Under optically thin conditions, \SiIV\ 1393.76 
$\angstrom$ is twice stronger than 1402.77 $\angstrom$, so the results we obtain in
this paper can also be applied to \SiIV\ 1393.76 $\angstrom$. Furthermore, the study of
\SiIV\ 1402.77 $\angstrom$ can provide theoretical support to our previous paper NS2017. 
In turn, the choice of the 1401.16 $\angstrom$ line for
oxygen is because the \OIV\ lines 
are faint and require longer exposure times to be observed \citep{De-Pontieu:2014vn}. 
Thus, in order to make any prediction that could be corroborated in future \IRIS\ 
analysis, we focus on the strongest of the oxygen lines, which has a better chance 
to be detected. For simplicity, hereafter we refer to the \SiIV\ 1402.77 $\angstrom$ 
and \OIV\ 1401.16 $\angstrom$ emissivities as the \SiIV\ and \OIV\ emissivities, 
respectively.

\setcounter{table}{0}
\begin{deluxetable}{c|ccccc}
\caption{Relevant parameters for the studied emission lines}\label{table1}
\tablehead{
\colhead{Line} & \colhead{$T_{_{SE}}$ (K)} & \colhead{$g_u$} & \colhead{$g_l$} & \colhead{$f_{lu}$} & \colhead{$n_u/\epsilon_{\lambda}$}}
\startdata
\SiIV\ 1402.77 \AA & $10^{4.9}$ & $2$ & $2$ & $2.7\times10^{-1}$ & $974$\\
\hline
\OIV\ 1401.16  \AA & $10^{5.2}$ & $6$ & $4$ & $5.1\times10^{-7}$ & $7.62\times10^{8}$ \\
\enddata
\end{deluxetable}

\Needspace{5\baselineskip} 
\subsubsection{Boundary conditions}
We are imposing periodicity at the side boundaries; 
for the vertical direction, characteristic conditions are implemented at the
top, whereas an open boundary is maintained at the bottom keeping a fixed value
of the entropy of the incoming plasma. Additionally, in order to
produce flux emergence, we inject a twisted magnetic tube through the bottom
boundary following the method described by \cite{Martinez-Sykora:2008aa}. 
The parameters of the tube (specifically, the initial location of
the axis, $x_0$ and $z_0$; the field strength there, $B_0$; the tube radius
$R_0$; and the amount of field line twist $q$) are identical in both
experiments and given in Table \ref{table2}. The total axial magnetic flux is
$\Phi_0 = 6.3 \times 10^{18}$~Mx, which is in the range of  
an ephemeral active region \citep{Zwaan:1987yf}. Details about this kind of
setup are provided in the paper by NS2016.

\begin{table}[h!]
\renewcommand{\thetable}{\arabic{table}}
\centering
\caption{Parameters of the initial twisted magnetic tube for both experiments} \label{table2}
\begin{tabular}{ccccc}
\tablewidth{0pt}
\hline
\hline
$x_0$ (Mm) & $z_0$ (Mm) & $R_0$ (Mm) & $q$ (Mm$^{-1}$) & $B_0$ (kG) \\
\hline
\decimals
15.0 & -2.8 & 0.10 & 2.4 & 20     \\
\hline
\end{tabular}
\end{table}

\begin{figure*}
\epsscale{1.18}
\plotone{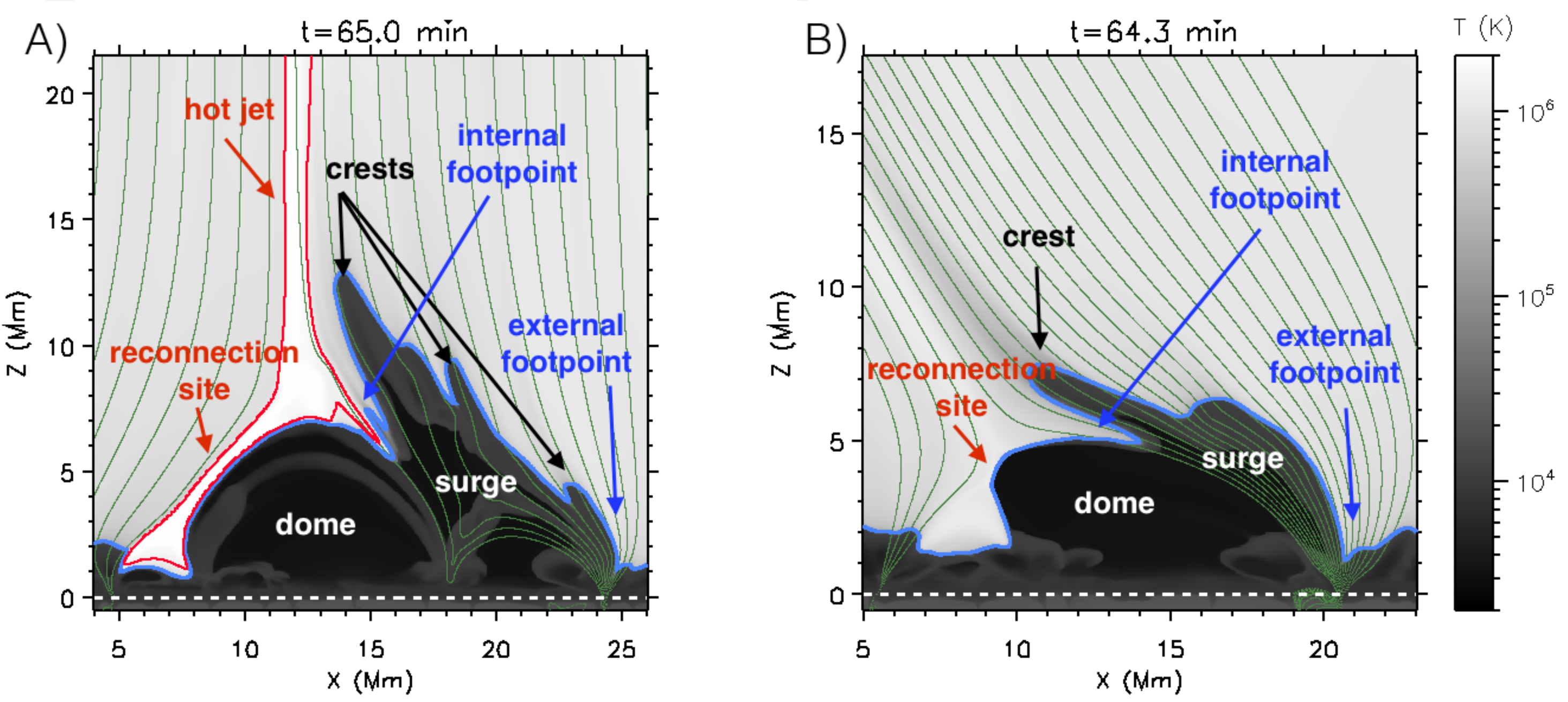}
\caption{Image (taken from NS2017) showing 2D 
temperature maps for the context of the surge experiments and the regions of interest. 
A) The vertical coronal magnetic field experiment at $t=65.0$ minute. 
B) The slanted coronal magnetic case 
at $t=64.3$ minute. Additionally, magnetic field lines (green), 
temperature contours for the $T_{_{SE}}=7.9 \times 10^{4}$ of \SiIV\ K (blue), and 
for $T=1.2 \times 10^{6}$ K (red) are added. \label{figure2}}
\end{figure*}

%
%
\Needspace{5\baselineskip}
\section{General features of the time evolution of the experiments}\label{sec:3}
The numerical experiments start with the injection of the twisted magnetic tube 
through the bottom boundary ($t=0$ minute). Within the convection zone, the tube rises
with velocities of $\lesssim 2$ km s$^{-1}$ and suffers deformations due to the
convection flows, mainly in the regions where the downflows are located. The twisted tube 
continues rising until it reaches the surface. There, the magnetized plasma accumulates
until it develops a buoyancy instability ($t \approx 40$ minute) in a similar way as
explained by NS2016.

The subsequent phases of evolution are characterized by the emergence and 
expansion of the magnetized plasma into the solar atmosphere, producing a dome-like 
structure of cool and dense matter ($t \sim 50$ minute). During the 
expansion process, the dome interior becomes rarefied due to gravitational flows. 
Simultaneously, the magnetic field of the emerged plasma collides with the preexisting 
coronal ambient field and, as a consequence, non-stationary magnetic reconnection occurs,
forming and ejecting several plasmoids. Our vertical experiment 
has recently been used by \cite{Rouppe2017} to show that the \SiIV\ spectral synthesis of those
plasmoids is able to reproduce the highly broadened line profiles,
often with non-Gaussian and triangular shapes, seen in \IRIS\ observations.

As an indirect consequence of the magnetic reconnection, a surge is obtained
in both experiments. This is illustrated in Figure \ref{figure2} through temperature 
maps with overlying magnetic field lines for each experiment: panel A, the vertical 
experiment at $t=65.0$ minute, and panel B, the slanted one at $t=64.3$ minute. Those
are representative instants when the surge is clearly distinguishable as an elongated 
structure detached from the dome. For later
reference, different regions have been marked in the figure that
  will be seen below to correspond to prominent features of the surge
in terms of NEQ ionization and brightness in the spectra: the internal footpoint, which is located at the base of a wedge created by the detachment process that separates the surge from the dome (NS2016); the external footpoint, which is
just the external boundary of the surge, and the flanks and top of the crests.
Although not directly discussed in this paper, another region is probably worth mentioning, namely the 
hot jet: in the vertical experiment, it is shown clearly through the red temperature contour of 
$T=1.2 \times 10^{6}$ K; in the slanted one, the temperatures of the high-speed collimated ejection 
are not distinguishable from the rest of the corona. The difference between both
experiments may be due to the fact that the slanted case has a denser emerged
dome and, perhaps, the entropy sources in it are less efficient in 
heating the plasma that passes through the magnetic reconnection site.

%
%
\Needspace{5\baselineskip}
\section{The role of the nonequilibrium (NEQ) ionization}\label{sec:4}
The importance of the NEQ ionization is studied in this section from a triple
perspective: (a) the comparison of the NEQ number densities with those calculated under the SE approximation
(Section~\ref{sec:4.1}); (b) the consequences for the emissivity of the
plasma (Section~\ref{sec:4.2}); and (c) the key mechanisms that cause the
departure from statistical equilibrium in the surge plasma
(Section~\ref{sec:4.3}).

%
%
\Needspace{5\baselineskip}
\subsection{The SE and NEQ number densities}\label{sec:4.1}
The results of the current paper are obtained by solving the equation rates for
the relevant ionization states of Si and O, i.e., taking into account
nonequilibrium effects using the \citet{olluri:2013aa} module mentioned in
earlier sections: the number densities of emitters $n_u$ thus calculated will be
indicated with the symbol $n_{_{NEQ}}$. In order to test the accuracy of the SE
approximation, we have also calculated the $n_u$ that would be obtained 
imposing statistical equilibrium in the \citet{olluri:2013aa} module: those will be 
indicated with the symbol $n_{_{SE}}$. The accuracy or otherwise of the SE 
approximation is measured here through the following ratio:

\begin{eqnarray}
	r & = &  \frac{n_{_{SE}} - n_{_{NEQ}}} { n_{_{SE}} + n_{_{NEQ}}  },
\label{eq:ratio}
\end{eqnarray}
The parameter $r$ varies between -1 and 1; its meaning is as follows:

\begin{enumerate}[a)]

\item If $r \approxeq 0$, the number density of emitters obtained imposing SE would be approximately
equal to the one allowing NEQ rates ($n_{_{SE}} \approxeq n_{_{NEQ}}$), so in those regions the SE approximation to calculate the state of ionization would be valid.

\item If clearly $r < 0$, this means that $n_{_{SE}} < n_{_{NEQ}}$, so the approximation of SE ionization would underestimate the real 
population. As $r$ becomes more negative, the NEQ
effects would be more prominent and the SE approximation 
would become less accurate. In the extreme case ($r = -1$), the assumption of SE  would mistakenly result in an absence of ions in the ionization state of interest!

\item On the other hand, if $r > 0$,  it follows that $n_{_{SE}} > n_{_{NEQ}}$,  so the computation 
of the ionization in SE would be wrong again, but in this case because it would overestimate the real population. When $r = 1$, SE would give as a result a totally fictitious population, since
the full NEQ calculation indicates that there are no ions!

\end{enumerate}

\begin{figure}
\epsscale{1.05}
\plotone{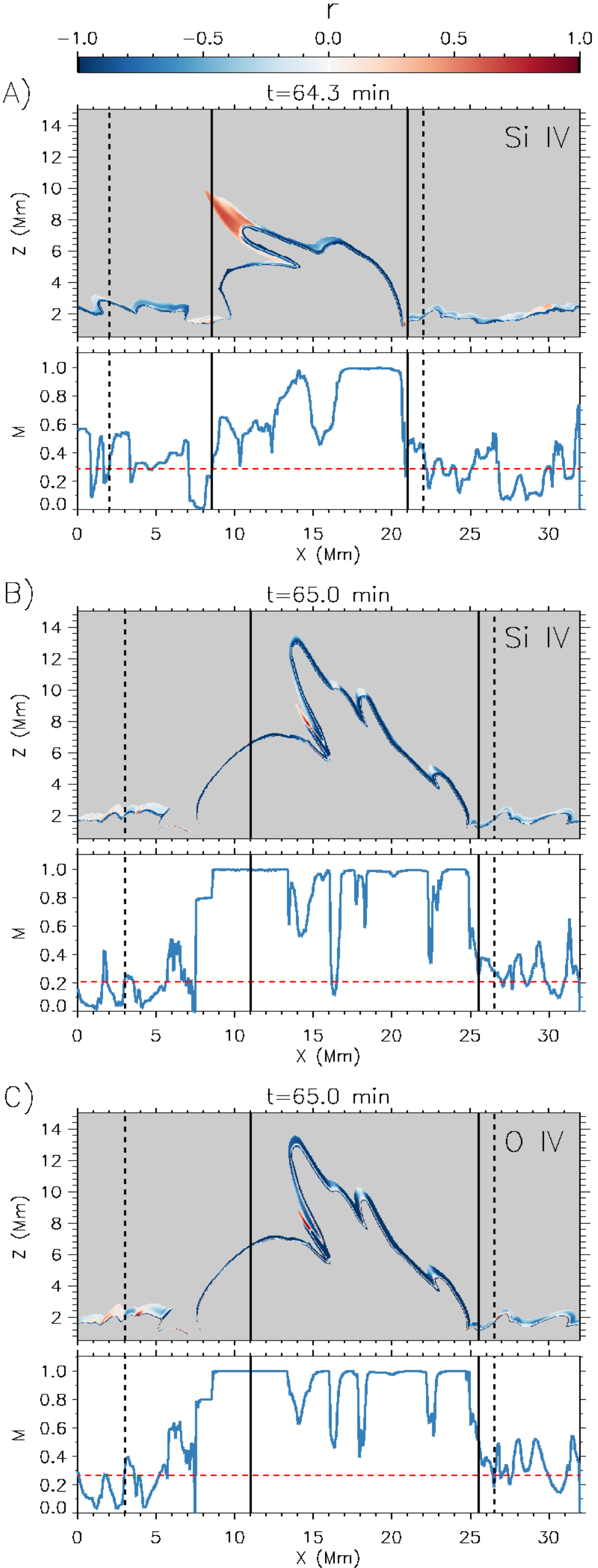}
\caption{2D maps of the ratio $r$ from Equation (\ref{eq:ratio}) for the
  \SiIV\ population for (A) the slanted experiment at $t=64.3$ minute and (B)
  the vertical one at $t=65.0$ minute. Panel C shows $r$ for the
  \OIV\ population in the vertical experiment at $t=65.0$ minute. 
  A gray color mask is overplotted where the emissivity obtained from
  the NEQ computation is $\epsilon_{_{CGS}} < 10^{-10}$. Below each 2D map,
  the median M of $\vert r \vert$ in the high-emissivity region (i.e., outside
  of the mask) is shown. Solid and dashed
  vertical lines delimit the ETR and QTR regions, respectively. The
  horizontal line in the M panels marks the average value of M within the
  QTR. \label{figure3}}
\end{figure}

The ratio $r$ is plotted in Figure~\ref{figure3} for
  the two experiments described in this paper. The upper panels in each block
  contain 2D maps of $r$, namely, for (A) \SiIV\ in the slanted experiment at
  $t=64.3$ minute; (B) \SiIV\ in the vertical experiment at $t=65$ minute;
  and (C) \OIV\ also in the vertical experiment at $t=65$ minute. To limit
  the diagram to the relevant regions, a grey mask is overplotted and only those pixels with emissivity obtained from
  the NEQ computation above a threshold ($\epsilon_{_{CGS}} \geq 10^{-10}$) are being shown. 
 The bottom panel in each block contains a line plot for the median M of the absolute value of $r$
   in the regions not covered by the mask in each column. Using
  the absolute value of the ratio elucidates the areas where NEQ ionization is 
 important, either because SE underestimates ($r < 0$) or overestimates 
 ($r > 0$) the real number density of emitters $n_u$. In the figure, two regions can be clearly distinguished:

1. The \textit{Quiet Transition Region} (hereafter QTR). We define it as 
the transition region that has not been perturbed by the flux emergence and subsequent surge 
and/or jet phenomena. The horizontal extent of the QTR is marked in the figure with dashed vertical lines 
and corresponds to the region located between $0.0 \leq x \leq 2.0$ and $22.0 \leq x \leq 32.0$ Mm, 
for the slanted experiment; and $0.0 \leq x \leq 3.0$ and $26.5 \leq x \leq 32.0$ Mm, for the vertical one. 
In this domain, $r$ mostly shows negative values (blue color in the image) in a thin 
layer in the transition region ($z \sim 2$ Mm). The corresponding M value is on average 
between $0.2$ and $0.3$ (horizontal dashed line in red in the panels), which indicates that both \SiIV\ and \OIV\ 
suffer significant departures from statistical equilibrium.

2. The second region corresponds to the main result of this section: the emerged 
domain, namely, the dome and surge, are severely affected by the NEQ ionization both for silicon 
and oxygen (see the dark blue color which corresponds to $r\approx-1$, and corresponding M value close to 1). 
The value $r\approx-1$ is found either in cold regions with $T \sim 2 \times 10^4$ K and
also in hot domains ($T \sim 5 \times 10^5$ K). Also, on the left of 
the surge, we also find some regions where $r > 0$ (red), especially in the slanted experiment,
which indicates that the SE approximation is overestimating the real population. Since our main goal is 
to study the surge, we focus on its surroundings, and, in particular, on the domain marked in the figure with solid lines, i.e.,  $8.5 \leq x \leq 21.0$ Mm, for the slanted experiment, and $11.0 \leq x \leq 25.5$ Mm, for the vertical one. In the following, we refer to this range as the \textit{Enhanced Transition Region} (ETR). 
 In the associated 1D panels, we see that M shows larger values than in the QTR; in fact, the median reaches values close to one in many places of the ETR . There are some specific locations within the ETR where M shows substantially lower values, e.g., $x=16.2$ Mm or $x=18.1$ Mm in the B and C panels . Looking at the 2D panels, we realize that in those locations, part of the TR has $r$ close to zero  (white patch above the blue line).  Consequently, the median in that vertical columns decreases. Note, however, that even in those locations the M values in the ETR are larger than, or at least comparable to, the largest ones found in the QTR. This finding highlights the relevance of including the NEQ calculation for eruptive phenomena like surges, since without it, the calculated \SiIV\ and \OIV\ populations would be totally erroneous. This would translate into wrong emissivity values and therefore mistaken synthesis diagnostics.

\begin{figure*}
\epsscale{1.0}
\plotone{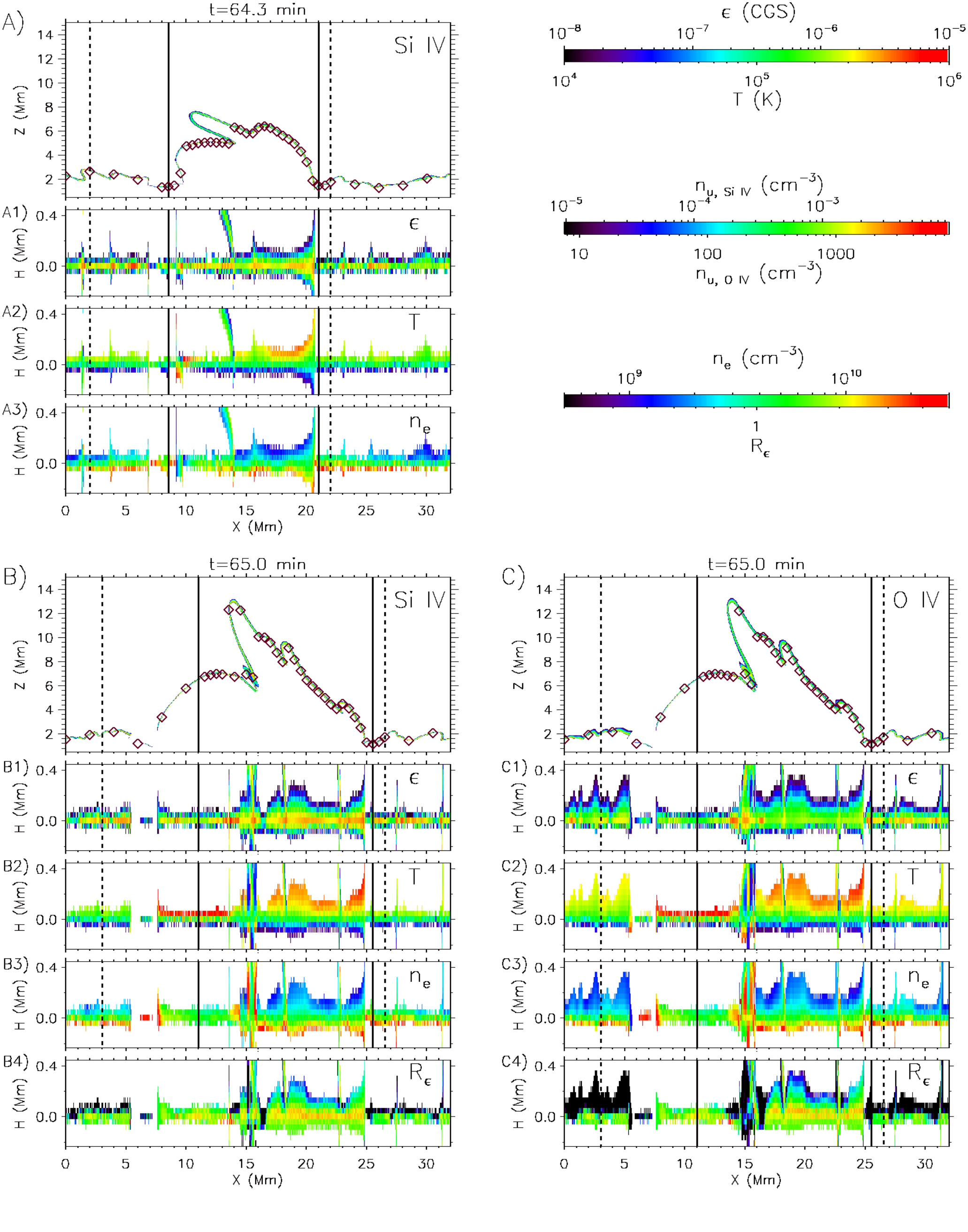}
\caption{
Maps of the 2D emissivity $\epsilon$ for A) \SiIV\ in the slanted experiment, 
B) \SiIV\ in the vertical one, and C) \OIV\ also for the vertical experiment. Diamonds have been superimposed on the region of high emissivity to mark the position of $z_{max}$ (see Equation \ref{eq:H}). 
A color scale at the right column of the figure contains the translation from emissivity to number densities of emitters $n_u$. In each of the blocks, secondary panels for $\epsilon$, $T$, and $n_e$ have been inserted that use $H$ as vertical scale. Additionally, a panel of the ratio of the \SiIV\ and \OIV\ emissivities, $R_{\epsilon}$, is added for B and C blocks. All the maps only show places where $\epsilon_{_{CGS}} \geq 10^{-8}$. The instants in the panels and the vertical lines are
the same as in Figure \ref{figure3}. The accompanying animation shows the time evolution of the
three experiments from the early stages of the surge until the its decay phase.\\
(An animation of the figure is available)}\label{figure4}
\end{figure*}

%
%
\Needspace{5\baselineskip}
\subsection{Characterizing the plasma in NEQ}\label{sec:4.2}
Once we have studied the NEQ effects on the two different domains of our 
experiments (ETR, QTR), we now turn to the associated question of the 
emissivity, in particular, for the \SiIV\ and \OIV\ lines. To that end, we start by showing 2D
maps of the emissivity in Figure \ref{figure4} (top panel in each block) for the same 
instants as in Figure \ref{figure3}. In this case, we have constrained the
maps to values of $\epsilon_{_{CGS}} > 10^{-8}$, just to 
focus on the layer with the largest emission, which is the natural candidate to be observed. 
Since the emitting layer is really thin, we are adding small 2D maps at the bottom of each block containing a blow-up of the emissivity, $\epsilon$, and, additionally, of the temperature, $T$; electronic number density, $n_e$; and the ratio between the \SiIV\ and \OIV\ emissivities, $R_{\epsilon}$. More precisely, for each vertical column we define a height coordinate $H$ centered at the position [called $z_{max}(x)$ in the following] of the 
maximum emissivity in that column:
\begin{eqnarray}
	H & = &   z - z_{max},
\label{eq:H}
\end{eqnarray}
and use it, instead of $z$, in the maps. For
  clarity, in the top panel we have indicated the location of $z_{max}$ at
  selected columns using symbols. Since emissivities can be converted into number densities of emitters $n_u$ via simple multiplication with a constant factor
  (Equation~\ref{eq:emissivity} and Table~\ref{table1}), a
  color bar with $n_u$ both for \SiIV\ and \OIV\ has been added in the figure.

  By comparing the two emissivity panels in each block of
  Figure \ref{figure4} (see also associated movie), we find that the region of high emissivity
  at the footpoints and crests of the surge covers a larger vertical range
  than in other regions. This is mainly  caused by the varying mutual angle of the vertical
  with the local tangent to the TR, so, in some sense, it is a line-of-sight
  (LOS) effect: full details of different LOS effects are discussed in Section
  \ref{sec:5.2}. Inspecting the lower panels of $\epsilon$ of each block, some locations (e.g.,
  the internal footpoint, $x \approx 15$ Mm, in \OIV\ at $t=65.0$ minute)
  are seen to have enhanced emissivity by a factor 2 or 3 in comparison to the
  maximum values usually seen at positions of the QTR and ETR; nonetheless, 
  this behavior is sporadic as seen in the accompanying movie. 
  
  We also notice that both \SiIV\ and \OIV\ show similar values of emissivity, in spite of the 
  huge contrast in the corresponding number density of emitters (see second 
  color scale at the right-top corner of the image). This is due to the difference in the 
  oscillator strengths $f_{lu}$, which for \OIV\ is six orders of magnitude weaker than for 
  \SiIV\ (see Table \ref{table1}). In panels B4 and C4, we have plotted 
  the emissivity ratio of \SiIV\ to \OIV, $R_{\epsilon}$, finding that the typical values 
  in the locations with the highest emissivity within the ETR are around 2 
  (although it can reach up to factors around 5), while in the QTR
  the average $R_{\epsilon}$ is close to 1. On the other hand, in the locations with low emissivity 
  and high temperature, especially in the 
  QTR, we appreciate that $R_{\epsilon}$ is lower than unity, which is not
  surprising since \OIV\ can be found at higher temperatures. Note
  that this is a ratio of emissivities and does not correspond to the intensity ratio commonly
  used for density diagnostics (e.g., \citealp{Hayes:1987,Feldman:2008,Polito2016}).

  We cannot find in the emissivity maps the same sort of drastic contrast 
  between QTR and ETR that we found for the $r$ parameter in the previous 
  section; nonetheless, we do appreciate differences between both
  regions in terms of temperature and electronic number density: the range of  $T$ and 
  $n_e$ in the ETR is larger than in the QTR. This is especially evident in the hot and 
  low-density part, where the $T$ and $n_e$ of the ETR reach values around 1 MK and 
  $10^9$ cm$^{-3}$, respectively (Note that the $n_e$ provided is obtained from local 
  thermodynamic equilibrium (LTE) since the ionization of the main contributors for electrons, such as 
  hydrogen and helium, are computed in LTE according to the equation-of-state table of Bifrost). 
  In order to further explore those differences, we
    resort to a statistical study of the values of emissivity and temperature
    in the different regions (QTR, ETR) and for the two ions, which is presented in the
    following. The statistics is based on all plasma elements with
    $\epsilon_{_{CGS}} \geq 10^{-8}$ in the time span between surge formation
    ($t = 55.0$ minute) and decay ($t = 70.7$ minute). The resulting sample
    contains $4 \times 10^6$ elements. Figure~\ref{figure5} shows the
    corresponding Joint Probability Density Functions (JPDFs) for 
    emissivity $\epsilon$ and temperature $T$ for the vertical
    experiment. For \SiIV\ we could also show results for the statistical distributions 
    for the slanted experiment, but the resulting JPDFs are very similar to those 
    presented here. This similarity suggests that,
    although the vertical and slanted experiments differ in 
    terms of magnetic configuration, size of the emerged dome, and shape of the 
    surge (compare the two panels of Figure \ref{figure2}), the results described below
    could be applicable to different surge scenarios. In the following, 
    we explain the results first for \SiIV\ (Section \ref{sec:4.2.1}), and then 
    for \OIV\ (Section \ref{sec:4.2.2}).

\Needspace{5\baselineskip}
\subsubsection{Plasma emitting in \SiIV}\label{sec:4.2.1}
\begin{figure}
\epsscale{1.18}
\plotone{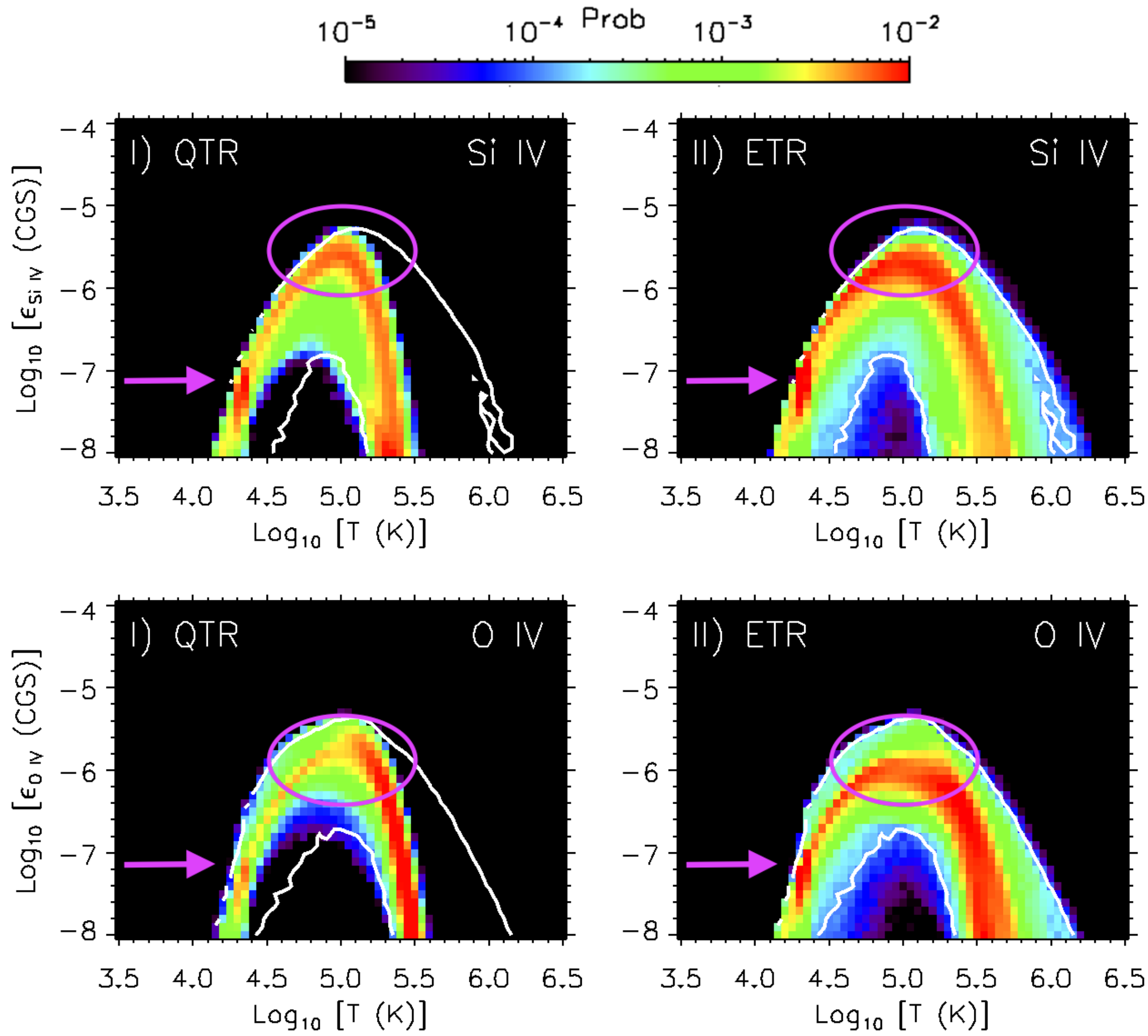}
\caption{JPDFs of emissivity and temperature in the QTR and ETR in
    the vertical experiment for \SiIV\ (top row) and \OIV\ (bottom row) for
    the time range $t=55.0$ minute to $t=70.7$ minute. The size of the sample is $4 \times
  10^6$ elements. The white lines are isocontours of probability
    equal to $10^{-4}$ in the ETR distribution. The areas
  marked by ovals and arrows are discussed in the text.\label{figure5}}
\end{figure}

We start analyzing the QTR and ETR distributions for the \SiIV\ emissivity
(see top row of Figure \ref{figure5}). The main result is that in the region
with the largest emissivity values the ETR is more densely populated than the
QTR (see the region marked by a pink oval around $\epsilon_{_{CGS}} \sim
10^{-5.6}$), i.e., the boundaries of the surges are more likely to show
signal in \SiIV\ observations than the QTR. This helps explain why, in the
\IRIS\ observations of our previous paper NS2017, we could detect the surge
as an intrinsically brighter structure than the rest of the TR. Furthermore,
both the QTR and ETR have the greatest values of emissivity in the
temperature range between $10^{5.0}$ and $10^{5.1}$ K. This differs from what
one would expect in SE, where the maximum emissivity is located at the peak
formation temperature ($T_{_{SE}} = 10^{4.9}$ K, see Table \ref{table1}),
again an indicator of the importance of taking into account NEQ effects. Additionally,
the distribution both for QTR and ETR is more spread in temperature than what one would expect from a
transition region distribution computed in SE (see, e.g., Figure 15 of the paper by \citealp{olluri:2015}). The
mass density found both for QTR and ETR around the maximum \SiIV\ emissivity
is $\rho \sim 6.3 \times 10^{-15}$ g cm$^{-3}$.

As part of the analysis, we have also found other features worth mentioning:

\begin{figure*}
\epsscale{1.18}
\plotone{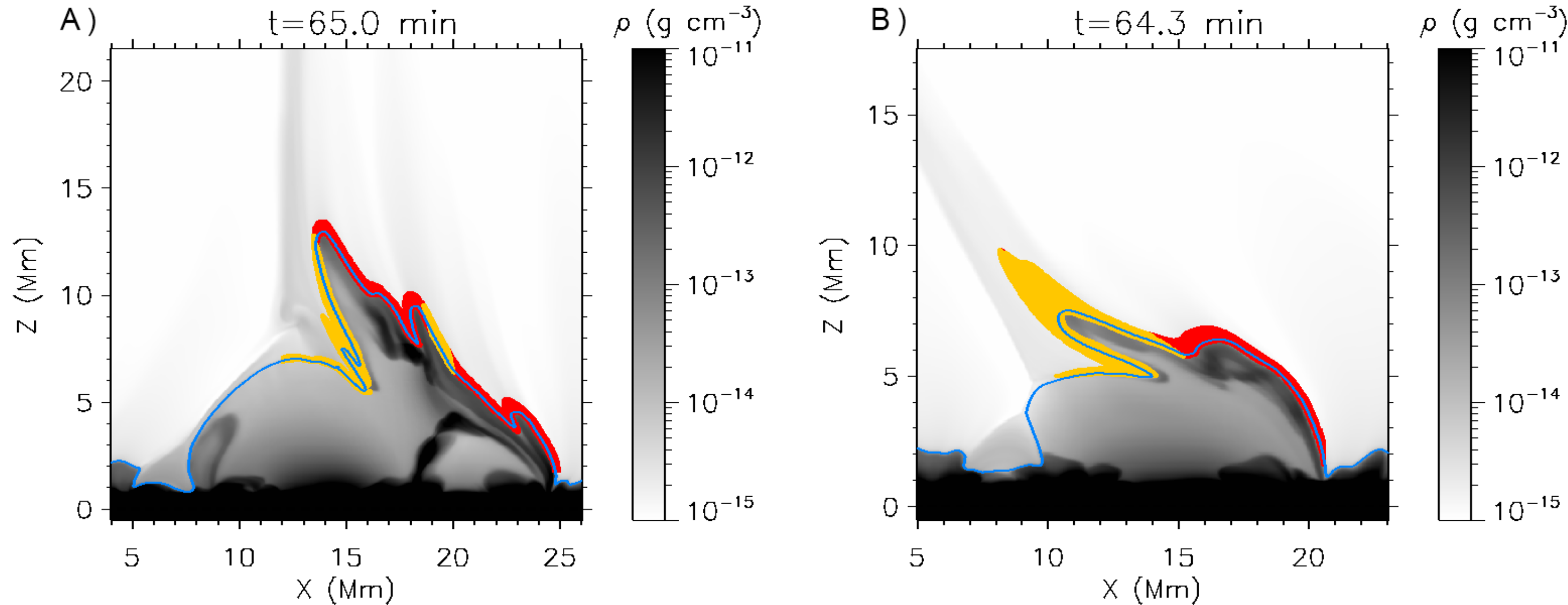}
\caption{Density map showing the basic distribution of the (approximately) 6000 Lagrange 
tracers used in the text, distributed into two parts drawn with yellow and
red dots corresponding to the two populations discussed in
Section~\ref{sec:4.3}. The accompanying animation shows the evolution of the
Lagrange tracers in the two experiments from  
early stages of the surge formation ($t \approx 55.0$ minute) until the its decay 
phase ($t \approx 70.0$ minute).\\
(An animation of this figure is available.)}
\label{figure6}
\end{figure*}  

\begin{itemize}

\item The ETR has a broader temperature distribution than the QTR. 
In order to illustrate this fact, all the panels of Figure \ref{figure5} contain isolines 
in white for the probability $10^{-4}$ in the ETR. 
The comparison of those contours with the QTR distribution shows that
the ETR has a wider distribution
in temperature, specially above $10^{5.5}$ K. Although not shown in the figure,  the mass density values for 
 most of the emitting plasma (more precisely: the mass density values with probability
 above $10^{-4}$) are constrained to similar ranges 
 for both the QTR and ETR: approximately at [$2.0 \times 10^{-15}$, $7.9 \times 10^{-14}$] g cm$^{-3}$, 
 for the vertical experiment; and at [$1.0 \times 10^{-15}$, $6 .0 \times 10^{-14}$] g cm$^{-3}$, for the slanted one.

\item A secondary probabilty maximum is located $\epsilon_{_{CGS}} \sim 10^{-7.2}$ and 
T $\sim 10^{4.3}$ K (see the arrows
in the panels). This corresponds to the temperature of the second ionization of helium 
according to the LTE equation-of-state of Bifrost:  the energy deposited
in the plasma is used to ionize the element instead of heating the plasma.
Including the NEQ ionization of helium should scatter the density probability in temperatures, as shown by 
\cite{golding:2016} in the TR of their numerical experiments
(the equivalent to our QTR); nevertheless, the NEQ computation of helium and 
a detailed discussion of their effects are out of scope of this paper. 

\end{itemize}

\Needspace{5\baselineskip}
\subsubsection{Plasma emitting in \OIV}\label{sec:4.2.2}
Focusing now on the statistical properties of the \OIV\ emission (see bottom row 
of Figure \ref{figure5}), we see that, like for \SiIV, the probability distributions for 
both QTR and ETR  differ from what we could expect for a SE distribution, 
since they are centered at temperatures between $10^{4.9}$ and $10^{5.0}$ K
instead of $T_{_{SE}} = 10^{5.2}$ K (see Table \ref{table1}). Furthermore,
also here, the ETR and QTR distributions are broader in temperature than what one would 
expect from SE. Further noteworthy features of the plasma emitting in \OIV\ are:

\begin{itemize}

\item The QTR exhibits larger probability than the ETR ($> 10^{-3}$) in
the maximum values of the emissivities ($\epsilon_{_{CGS}} = 10^{-5.8}$); 
nevertheless, this fact changes around $\epsilon_{_{CGS}} = 10^{-6.0}$,
where the ETR shows larger emissivity (compare the region within the colored oval). 
Due to this complex behavior, we need to integrate the emissivity to know whether the 
ETR can be detected as a brighter structure compared to the QTR. In Section  
\ref{sec:5} we discuss this fact analyzing the obtained synthetic profiles.

\item  The ETR shows emissivity in \OIV\ in a larger range of temperatures than 
the QTR, which is akin to the result for \SiIV\ described in Section
\ref{sec:4.2.1}. This difference in the ranges is apparent mainly in hot coronal
 temperatures comparing the probability contours of the ETR (in white) 
 with the QTR distribution.

\item We find the same secondary probability maximum as in the \SiIV\
panels at the temperature of the second ionization of helium (see the pinks arrows).

\item Comparing the \OIV\ panels with the \SiIV\ ones, we see that the \OIV\ 
distribution is more populated in hot temperatures, and correspondingly, 
lower densities. This is something we could expect since the ionization of this particular
oxygen ion occurs at higher temperatures than \SiIV.

\end{itemize}

%
%
\Needspace{5\baselineskip}
\subsection{Lagrange tracing: how the entropy sources affect the NEQ ionization}\label{sec:4.3}
We focus now on the role of the entropy sources in the emissivity and 
NEQ ionization. To that end, we follow in time $ \approx 6000$ plasma elements of the 
ETR through Lagrange tracing. In the following, we explain the set up for the Lagrange elements
(Section \ref{sec:4.3.1}) and the results obtained from their tracing (Section \ref{sec:4.3.2}).

\subsubsection{The choice of the Lagrange elements}\label{sec:4.3.1}

The Lagrange elements are selected at a given instant, corresponding to an
  intermediate evolutionary stage when the surge is clearly 
  distinguishable as a separate structure from the dome. The
    selected instants are $t=64.3$ minute for the slanted experiment and
    $t=65.0$ minute for the vertical experiment, which are the same times
    used for Figures \ref{figure2}, \ref{figure3}, and \ref{figure4}. In order to focus on 
    the domain in and near the surge, we limit the selection to
  the rectangular areas: $12.0 \leq x \leq 25.2$, $2.2 \leq z \leq
  15.0$ (vertical experiment); and $7.0 \leq x \leq 19.0$, $5.0 \leq z \leq
  15.0$, and $19.0 \leq x \leq 20.5$, $1.5 \leq z \leq 15.0$ (slanted
  experiment). On those rectangles we lay a grid with uniform spacing $\Delta
  x = \Delta z =40$~ km: the Lagrange elements are chosen among the
  pixels in that grid. A further criterion is then introduced: we are
  interested in studying the origin and evolution of the plasma elements with
  strong emission in \SiIV\ and \OIV. Thus, a lower bound in the \SiIV\ and
  \OIV\ emissivity is established, namely $\epsilon_{_{CGS}} > 10^{-10}$,
  discarding all the pixels with emissivities below that value at the
  instants mentioned in the previous bullet point. The resulting choice of
  Lagrange tracers is shown in Figure \ref{figure6} as red and yellow dots
  (the colors serve to distinguish the populations described below). Once the distribution is 
  set, we then follow the tracers backward in
  time for 10 minutes, to study their origin, and forward in time for 5.7
  minutes, to see the whole surge evolution until the decay phase, with a
  high temporal cadence of $0.2$~seconds (see the accompanying animation to
  Figure~\ref{figure6}).

\subsubsection{Plasma populations and role of the entropy sources}\label{sec:4.3.2}
\begin{figure}
\epsscale{1.15}
\plotone{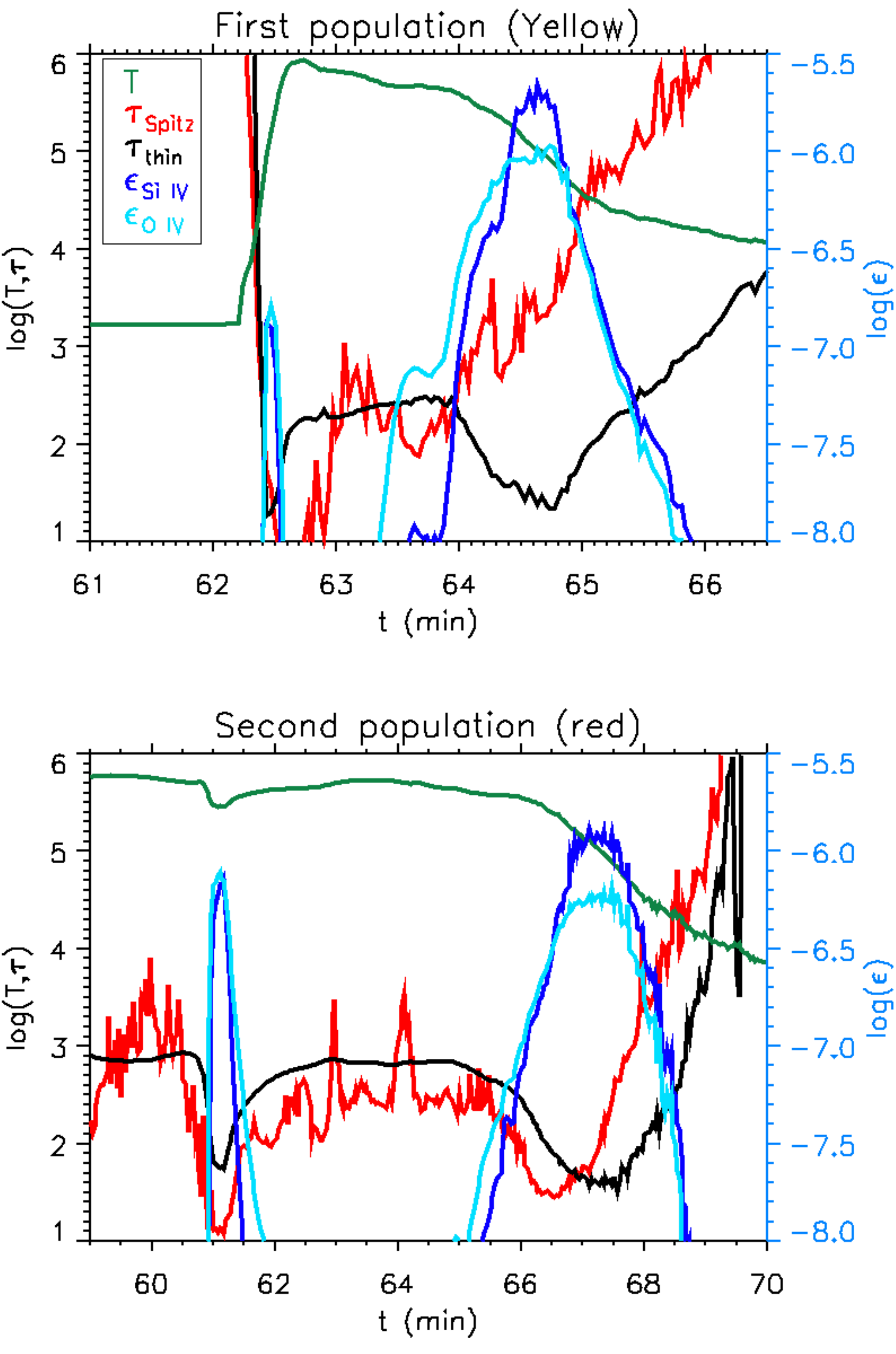}
\caption{Time evolution of key physical quantities for representative
  Lagrange tracers in the vertical experiment of Figure~\ref{figure5}.  Top:
  Lagrange element coming from the emerged dome (yellow population in
  Figure~\ref{figure6}). Bottom: Lagrange originating in the corona (red
  population in Figure~\ref{figure6}).  The curves show (left
    ordinate axis) the logarithm of temperature $T$ (green); of the
    characteristic time of the optically thin losses $\tau_{thin}$ (black);
    and of the characteristic time for the thermal conduction $\tau_{Spitz}$
    (red); and (right ordinate axis) the logarithm of the \SiIV\ emissivity
    (dark blue); and of the \OIV\ emissivity (light blue). All quantities are
    in CGS units. \label{figure7}}
\end{figure}

Studying the time evolution of the Lagrange tracers, in particular their
thermal properties, one can distinguish two populations that are the source of 
the \SiIV\ and \OIV\ emission: one originating in the emerged dome (yellow plasma population in
Figure~\ref{figure6}) and the other one originating in the corona (red
population). By carefully inspecting the tracers of each population, we find that
their behavior is well defined: the major difference between the elements within
the same population is not the nature or order of the physical events described below, 
but rather the starting time of the evolution for each tracer. Figure~\ref{figure7} contains the time evolution of
different quantities as measured following a representative Lagrange element of
  each population, namely temperature, $T$, (green); \SiIV\ emissivity,
$\epsilon_{_{Si\ IV}}$, (dark blue); \OIV\ emissivity, $\epsilon_{_{O\ IV}}$,
(light blue); characteristic time of the optically thin losses,
$\tau_{thin}$, (black); and characteristic time for the thermal conduction,
$\tau_{Spitz}$ (red).

\textit{The first population} (top panel of Figure~\ref{figure7}, corresponding to the
  elements marked in yellow in Figure~\ref{figure6}), starts as
    cool and dense plasma coming from the emerged dome with extremely low
    emissivity (see the curves for the temperature in green, and for the
    emissivities in dark and light blue). At some point
  that plasma approaches the reconnection site and passes through the current
  sheet, thereby suffering strong Joule and viscous heating and
    quickly reaching TR temperatures. The sharp
    spike in the \SiIV\ and \OIV\ emissivity (blue curves) around $t\sim
    62.5$ minute corresponds to this phase: the temperature increase leads to
    the appearance of those ionic species, but, as the plasma continues being
    heated, it reaches high temperatures (maximum around $10^6$ K) and the
    number densities $n_u$ of \SiIV\ and \OIV\ decrease again. At those high
  temperatures the entropy sinks become efficient, with short
  characteristic times: see the red ($\tau_{Spitz}$) and black
  ($\tau_{thin}$) curves. The plasma thus enters a phase of gradual
    cooling, going again through TR temperatures, renewed
    formation of the \SiIV\ and \OIV\ ions, and increase in the corresponding
    emissivity (broad maximum in the blue curves in the right half of the
    panel). The plasma elements, finally, cool down to chromospheric
  temperatures, with the emissivity decreasing again to very low
  values.

 The defining feature of \textit{the second population}  (bottom panel of
   Figure~\ref{figure7}, red dots in Figure~\ref{figure6}) is that it
   originates in the corona as apparent in the temperature curve
   (green). This population starts at heights far above the reconnection
   site, with standard coronal temperature and density. During the magnetic
   reconnection process, its associated field line changes connectivity,
   becoming attached to the cool emerged region. Consequently, a steep
   temperature gradient arises along the field line, so the thermal
   conduction starts to cool down the plasma; given the temperature range,
   also the optically thin losses contribute to the cooling, although to a
   lesser extent (see the $\tau_{Spitz}$ and $\tau_{thin}$ curves around
   $t\sim 61$ minute, in red and black, respectively). The temperature drops
   to values around T $\sim 10^{5.5}$ K, which, according to the
   JPDFs of Figure~\ref{figure5}, makes it sufficiently likely that the
   \SiIV\ and \OIV\ emissivities from the Lagrange element are high. This
   explains the large increase, by a few orders of magnitude, in the blue
   emissivity curves around $t\sim 61$ minute 
   (although a small factor  $\sim 4$  is due to the simultaneous 
   increase in the mass density, which is reflected in a linear fashion in the emissivity). 
   This cooling to TR
   temperatures, however, is short lived: as the plasma element itself passes
   near the current sheet, it can
   be heated because of the Joule and viscous terms and the temperature
   climbs again to values where the emissivities are low: hence the sharp
   spike in the blue curves between $t\sim 61$ and $t\sim 61.5$ minute. There
   ensues a phase of gradual cooling from $t \sim 64$ onward, similar to what
   happened to the previous population, with characteristic cooling times of
   a few to several hundred seconds (see red and black curves), passage
   through TR temperatures, broad maximum in the emissivity
   curves and with the plasma finally reaching chromospheric temperatures.

\begin{figure}
\epsscale{1.18}
\plotone{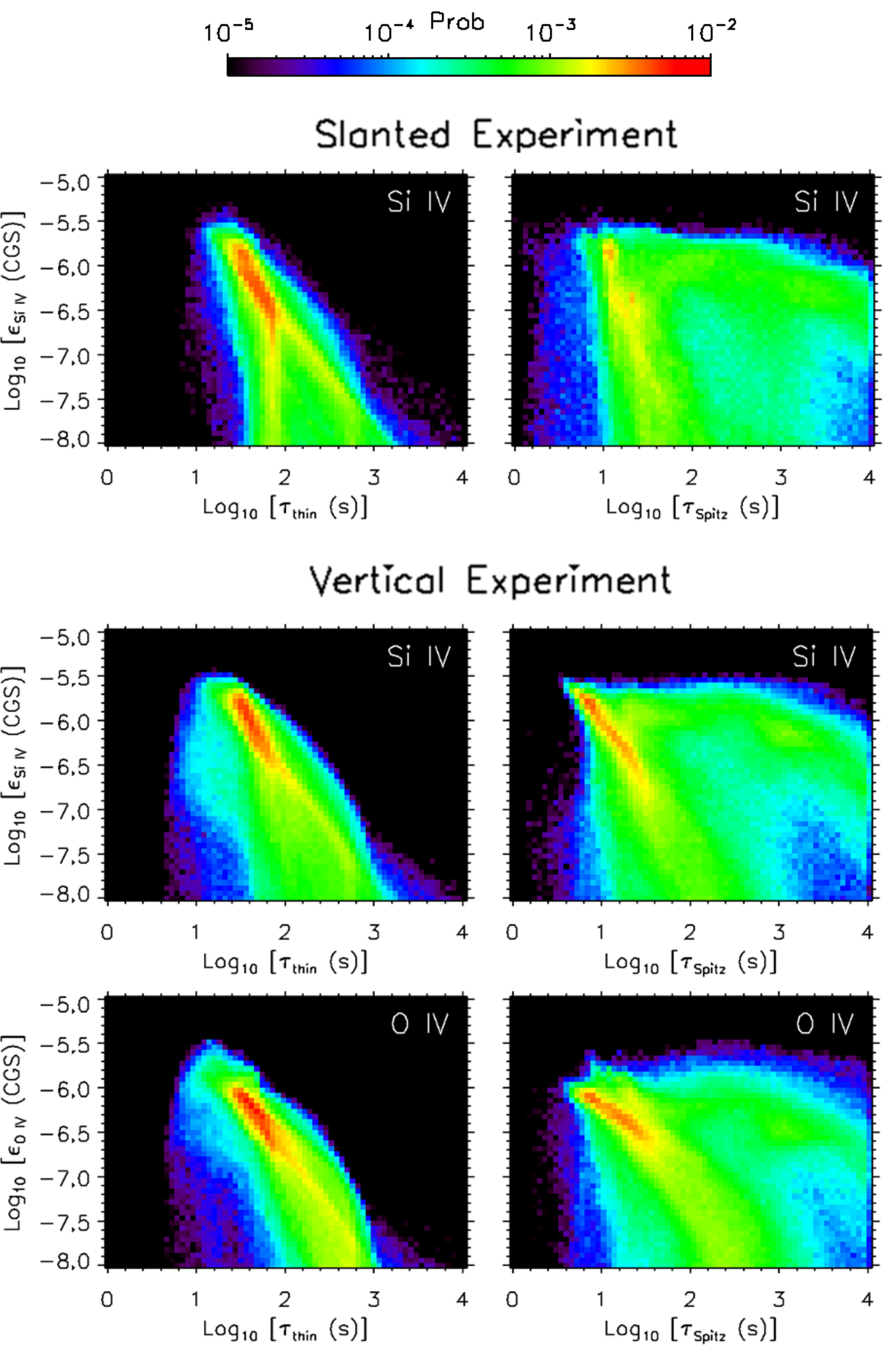}
\caption{
JPDFs for the \SiIV\ and \OIV\ emissivity of the Lagrange tracers
  over $15.7$ minutes versus the 
characteristic time of the optically thin losses ($\tau_{thin}$, left column) and the thermal conduction 
($\tau_{Spitz}$, right column). \label{figure8}}
\end{figure}

\begin{figure*}
\epsscale{1.21}
\plotone{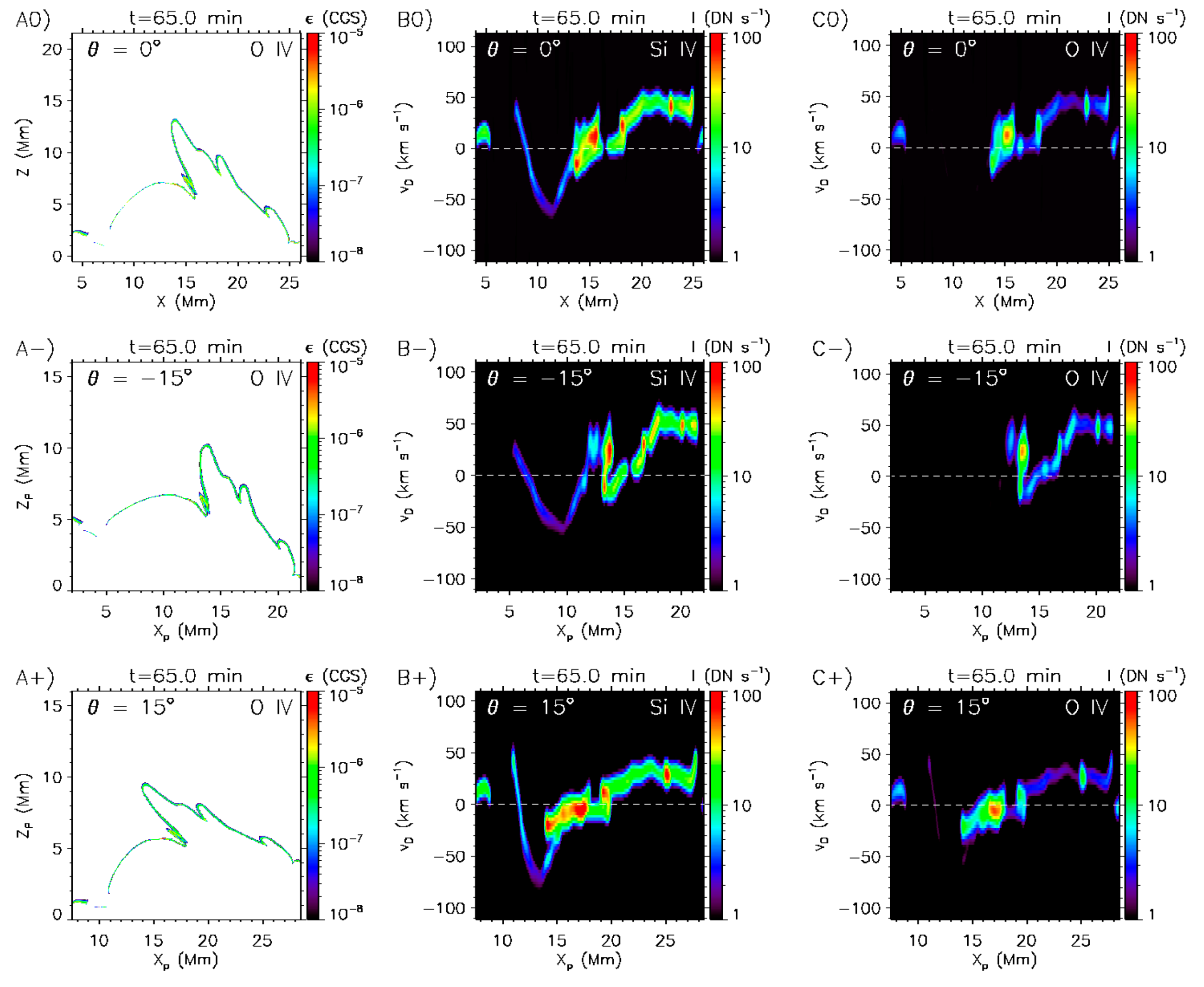}
\caption{Synthetic profiles for the vertical experiment at different times for various LOS ($\theta$). 
A) 2D map of the $\epsilon_{_{CGS}}$ for \OIV\ to show the context; B) synthetic spectral 
intensity for \SiIV, and ; C) synthetic spectral 
intensity for \OIV. To identify the LOS of each panel, we have added the symbols $0$, $-$, and
$+$ respectively to $\theta = 0, -15$ and $15$ \degree. In the $\theta \neq 0$ rows, $z_P$ and $x_P$ are, respectively, the vertical and horizontal 
coordinates of the rotated figures. The animation of this figure shows 
the time evolution of the surge from its origin ($t = 61$ minute) up to its decay phase 
($t = 70.7$ minute) in the vertical experiment for the three LOS.\\
(An animation of this figure is available.)} \label{figure9}
\end{figure*}   

In our previous paper (NS2016) we found that surges were constituted by four
different populations according to their thermal evolution. In the current
paper, we see that only two of them, labelled Populations B and D in the
NS2016 paper, are behind the enhanced emissivity of TR lines
like those from silicon and oxygen discussed here. The other two populations
described by NS2016 (A and C) keep cool chromospheric temperatures during
their evolution and do not play any role for the TR elements.

Using the Lagrange tracing method developed here, we can produce conclusive
evidence of enhanced \SiIV\ and \OIV\ emissivity and occurrence of fast evolution 
due to short-time scales in the entropy sources associated with heat conduction 
or optically thin radiative cooling. Figure~\ref{figure8} contains double PDFs for $\epsilon{_{CGS}}$
versus $\tau_{thin}$ (left panels) and $\tau_{Spitz}$ (right panels) using
as statistical sample the values of those quantities for all Lagrange
tracers along their evolution. The choice of the ionic species (\SiIV, \OIV) and
experiment (slanted, vertical) in the panels is as in Figures \ref{figure3} and \ref{figure4}.
The figure clearly shows that when the entropy sources act on short time
scales, the (\SiIV, \OIV) emissivities are large. In fact, the maximum
values of $\epsilon{_{CGS}}$ correspond to characteristic cooling times
between $20-100$~s for $\tau_{thin}$ and between $4-40$~s for
$\tau_{Spitz}$. Those
   changes are fast enough for the ionization levels of those elements to be
   far from statistical equilibrium.

%
%
\Needspace{5\baselineskip}
\section{Observational consequences}\label{sec:5}
In the paper by NS2017,  different observed \SiIV\ signatures
within the surge were analyzed. Moreover, counterparts to the observational features were
identified in the synthetic spectral profiles obtained from the numerical
model; however, a theoretical analysis to understand the origin of the
spectral features and the reason for the
brightness in the various regions of the surge was not addressed. In the
following subsection a theoretical 
study is carried out trying to quantify the impact of the NEQ ionization of silicon
and oxygen on the spectral and total intensities and the observational consequences thereof
(Section \ref{sec:5.1}). Then, given the involved geometry of the
  surge, the particular LOS for the (real or synthetic) observation turns out to be
crucial for the resulting total intensity and spectra. This is studied in Section \ref{sec:5.2}.

%
%
\Needspace{5\baselineskip}
\subsection{Synthetic profiles}\label{sec:5.1}
Figure \ref{figure9} contains the synthetic profiles obtained by integrating the 
emissivity along the line of sight for different wavelengths in the \SiIV\ $1402.77$ 
\AA\ and \OIV\ $1401.16$ \AA\ spectral region and for the vertical experiment. The 
three rows of the figure correspond to different inclination angles $\theta$ for the 
LOS: from top to bottom, 0\degree, -15\degree\ and 15\degree, respectively. The 
panels in each row contain A) the context of the experiment through a 2D map of 
the \OIV\ emissivity; B) the synthetic spectral intensity for \SiIV\ with the spectral 
dimension in ordinates and in the form of Doppler shifts from the central 
wavelength; and C) the corresponding synthetic spectral intensity for \OIV. 
Those spectra are obtained taking into account the Doppler shift due to the plasma 
velocity and applying a spatial and spectral PSF (Gaussian) degradation 
as explained in detail by \cite{Martinez-Sykora:2016obs}, their section 3.1, and by 
NS2017, their section 2.2. In this way, we will be able to directly
compare the results with \textit{IRIS}
observations. In order to ease the identification of the LOS in each panel, 
we have added to the labels on it, the symbols $0$, $-$, and $+$ respectively to $\theta = 0, -15$ and $15$ \degree. In the middle and bottom rows of the image, $x_P$ and 
$z_P$ are, respectively, the horizontal and vertical coordinates of the rotated figures. 

\begin{figure}
\epsscale{1.21}
\plotone{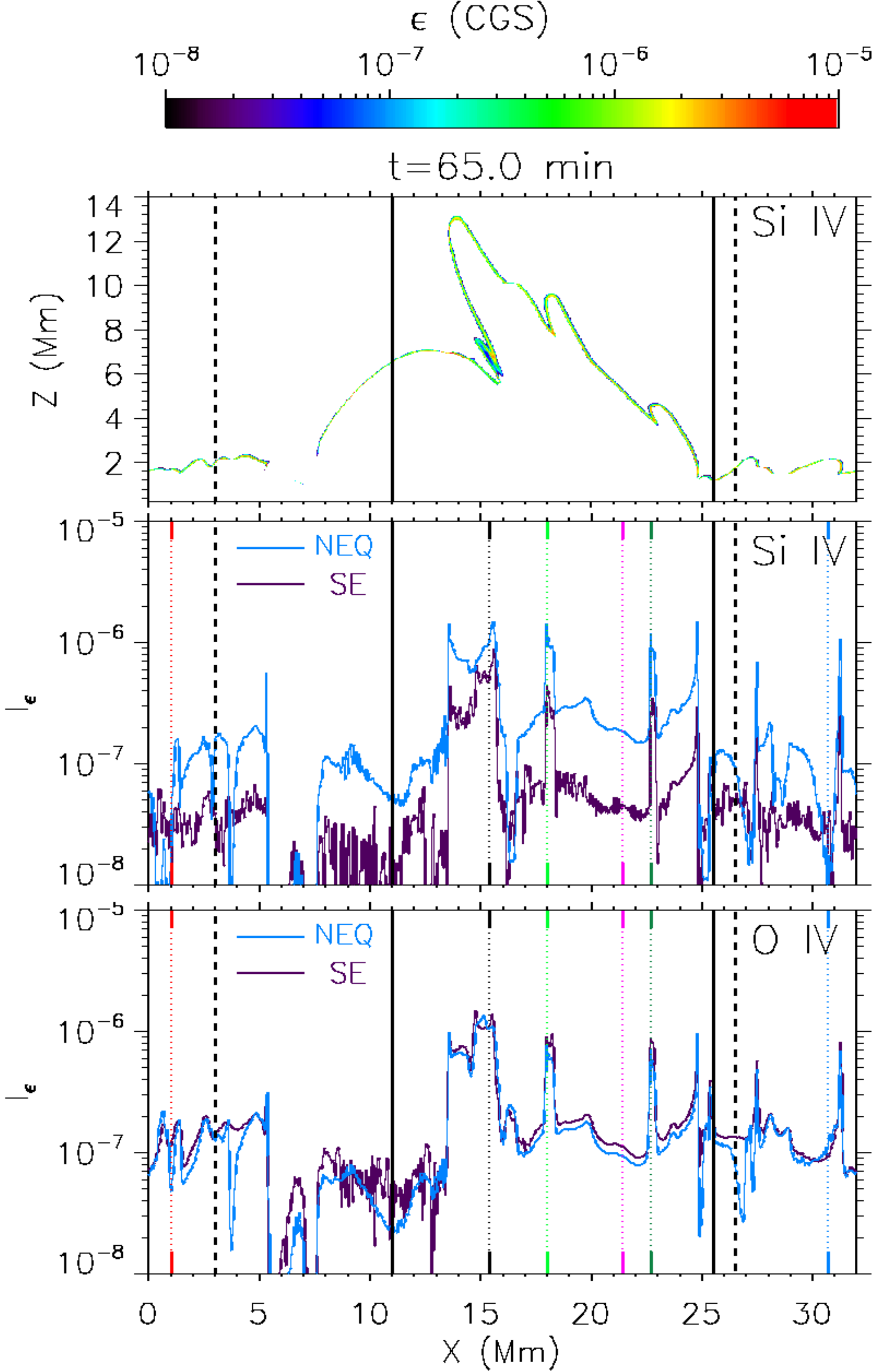}
\caption{Top panel: 2D map of the \SiIV\ emissivity. The
vertically integrated intensity $I_{\epsilon}$ is shown both for \SiIV\ (middle panel) and
\OIV\ (bottom panel) assuming NEQ ionization (light blue curve) and SE (purple).
Solid and dashed lines are overplotted in the image to 
delimit the ETR and QTR regions as in previous figures. Dotted vertical lines are
superimposed in the middle and bottom panel correspondingly to the cuts shown
in Figure \ref{figure11}.} \label{figure10}
\end{figure}

To extend the analysis, it is also of interest to
consider the total emitted intensity for each vertical column, i.e.,
\begin{eqnarray}
	I_{\epsilon} (x) & = &   \int_{z_0}^{z_f}{ \epsilon\; dz}\;,
\label{eq:ie}
\end{eqnarray}
which, following Equation \ref{eq:emissivity}, is equal to the column density 
of the emitting species along the LOS except for a constant factor). Equation (\ref{eq:ie})
has been calculated separately for \SiIV\ and \OIV\ and with the
emissivities obtained assuming either NEQ or SE ionization, to
better gauge the importance of disregarding the NEQ 
effects. The results are shown in the middle and bottom
panel of Figure \ref{figure10} for \SiIV\ and \OIV\ , respectively.  The top
panel of the image contains the 2D map of the emissivity for \SiIV\ for
context identification.  Combining Figures \ref{figure9} and \ref{figure10},
we are able to discern and describe characteristic features of the spectral
profiles.

\begin{itemize}

\item 
The most prominent feature in the synthetic profiles of Figure \ref{figure9}
is the brightening associated with the location of the internal footpoint of
the surge. In the corresponding movie, we can see how that footpoint is
formed at around $t=64$ minute, as the surge detaches from the emerged
dome. During those instants, the associated synthetic profiles are
characterized by large intermittent intensities and
bidirectional behavior with velocities of tens of km s$^{-1}$, as apparent,
e.g., in the B0 and C0 panels at $x \in [15,16]$ Mm. In
\SiIV\ and \OIV\ (B and C panels), we find that the internal footpoint is
usually the brightest region, although there are some instants in which the
brightest points can be located in the crests or the external footpoint. This
is a potentially important result from the observational point of view
because it can help us to unravel the spatial geometry of the surge in future
observational studies: if strong brightenings are detected in \SiIV\ and also
in \OIV\ within the surge, it would be reasonable to think that they
correspond to the internal footpoint of the surge. In this region, 
the intensity ratio between \SiIV\ and \OIV\ ranges between 2 and 7, approximately. Note
that, in general, the intensity ratio values vary depending on the observed region 
and features \citep{Martinez-Sykora:2016obs}.

In Section \ref{sec:5.2} we will see that LOS effects play a major role in causing 
the large brightness of the internal footpoint (and other bright features) compared to 
the rest of the surge. Here we consider the parallel question of the role of NEQ: what would be obtained for the intensity of the internal footpoint if SE were assumed? Comparing the values for \SiIV\ in Figure
\ref{figure10} (middle panel, $13.8 \le x \le 16.1$), there is roughly a
factor $2$, in the average, between the NEQ and SE intensities; for \OIV\ (bottom
panel), there is no major difference in the intensity between both
calculations. One could conclude that while NEQ is important for
  the \SiIV\ diagnostics, SE could be applied for the
  \OIV\ case; nonetheless, even in the latter case, although the
  NEQ and SE intensities are similar, one would make a mistake in the
  determination of derived quantities like the number densities of emitters (Section
\ref{sec:4.1}) and temperatures (Section \ref{sec:4.2}).

Further distinctive brightenings in the spectral profiles appear at the site 
of the crests and of the external footpoint of the surge. The brightness of those
  regions is clear in the \SiIV\ profiles (B0, B$-$ and B$+$
panels of Figure \ref{figure9}) through their large intensity, and is sometimes
comparable or greater than that of the internal footpoint (see, e.g., the
locations at $x \sim 13$, $18$, $23$ and $25$ Mm in the top row, or, $x \sim
14, 19$ and $25$ Mm in the bottom row). In the \OIV\ profiles (C0, C$-$ and C$+$ panels),
although faint in comparison with \SiIV\ (around a factor 5 less intense), 
most of those features are still slightly brighter than the rest of the surge. This difference in intensities
between \SiIV\ and \OIV\ can also be used to understand the observations: if
strong signals are observed in \SiIV\ associated with some moderate signal in
\OIV, it could indicate that we are detecting the crests or the external
footpoint. Concerning the NEQ/SE comparison of the intensity (Figure
\ref{figure10}), the crests and footpoints show the same behavior as the
internal footpoint.

\item The intensity of the rest of the surge is small in comparison
  with that of the footpoints and crests just described, so we
  wonder whether one could see it as a bright structure in actual
    observations and distinguish it from the rest of the TR.
  In the middle panel of Figure \ref{figure10}, comparing the $I_{\epsilon}$
  values for \SiIV\ within the ETR against those in the QTR, we can see that
  all the regions in the surge have a higher intensity than the
  QTR; outside of the brightest features, the excess emission of
    \SiIV\ in the surge may be just a factor 2 or 3 above the QTR, but that
  can provide enough contrast to tell the two regions apart
  observationally, as found in the NS2017 paper. Note, importantly,
   that there is a large difference between the NEQ and SE results
  for \SiIV\ in the surge, up to a factor 10, so
  the SE assumption would seriously underestimate the
  intensity. In fact, in most of the places \SiIV\ would be similar or fainter
   than \OIV\ if SE were valid as shown also, e.g. by \cite{Dudik:2014} and \cite{Dzifcakova:2017}.
  On the other hand, for \OIV\ (bottom panel), the prominent features are the footpoints and crests
  of the surge, while the other parts have $I_{\epsilon}$
  comparable to the QTR. As a consequence and from an observational point of view, while in \SiIV\ we
  could find enhanced emission in the whole surge (which is compatible with
  the NS2017 observation), for \OIV\ only the brightest regions would stand
  out, namely, the internal footpoints and, to a lesser extent, the crests and external footpoints. 
  
  The underlying reason for the enhanced brightness of all those features, 
  mainly the footpoints and crests of the surges, is not just the presence of  
  additional numbers of ions due to NEQ effects: the complex geometry of
  the surge TR has important consequences when integrating the emissivity
  along the line of sight to obtain intensities . This is discussed in the
  next section.

\end{itemize}

%
%
\Needspace{5\baselineskip}
\subsection{The role of the LOS}\label{sec:5.2}
The observation of
  TR lines generated in the surge strongly depends on the particular
  LOS. We show this here through two different effects:

\begin{enumerate}[a)]

\item The alignment of the LOS with respect to the orientation of the surge's
  TR.  We can appreciate this effect, e.g., in A0 panel of Figure \ref{figure9}. 
  There, considering,
  e.g., the crests situated at $x =13.5$, $x=18$ or $x=23$ Mm, we see that a
  vertical LOS grazing the left side of the crest will include contributions
  from a much larger length of the TR than if the crossing were perpendicular
  or nearly so. The same can be said of the external footpoint at $x=24.8$ Mm
  and also of the internal footpoint around $x = 16$ Mm.  This effect clearly
  enhances the brightenings seen in those values of $x$ in panels B0 and C0.
  Further evidence can be found by checking the $I_{\epsilon}$ curves in the
  middle and bottom panels of Figure \ref{figure10}; in fact, since
    the effect is purely geometrical, the contribution to brightness can be seen both in the
    NEQ and SE curves in the two panels of the figure. 
    Varying the angle of the LOS, we can reach enhancement factors 
    between 2 and 4; nonetheless, discerning which part of that factor is exclusively due to the 
    LOS is complicated, since variations in the angle of integration imply integrations along
    slightly different emitting layers. Additionally, the
  inclination of the LOS with respect to the surge's TR may be important for
  the apparent horizontal size of the brightenings. This can be seen through
  comparison of the three rows of Figure \ref{figure9}: considering the size
  of the brightening associated with the internal footpoint, we see that it
  covers a larger horizontal range in the $\theta=0$\degree\ and $\theta=15$\degree\ cases (top and bottom rows), than in the $\theta=-15$\degree\ case
  (middle row), since the good alignment of the latter is lost in the former.

\item The multiple crossings of the TR by individual lines of sight. 
Given that the TR of the surge is folded, there are horizontal ranges in
which the LOS crosses it more than once (typically three times, in some
limited ranges even five times). Given the optically thin
approximation, the emitted intensity in those lines of sight may be
a few, or several, times larger than the value in a single crossing. 

\end{enumerate}

\begin{figure}
\epsscale{1.21}
\plotone{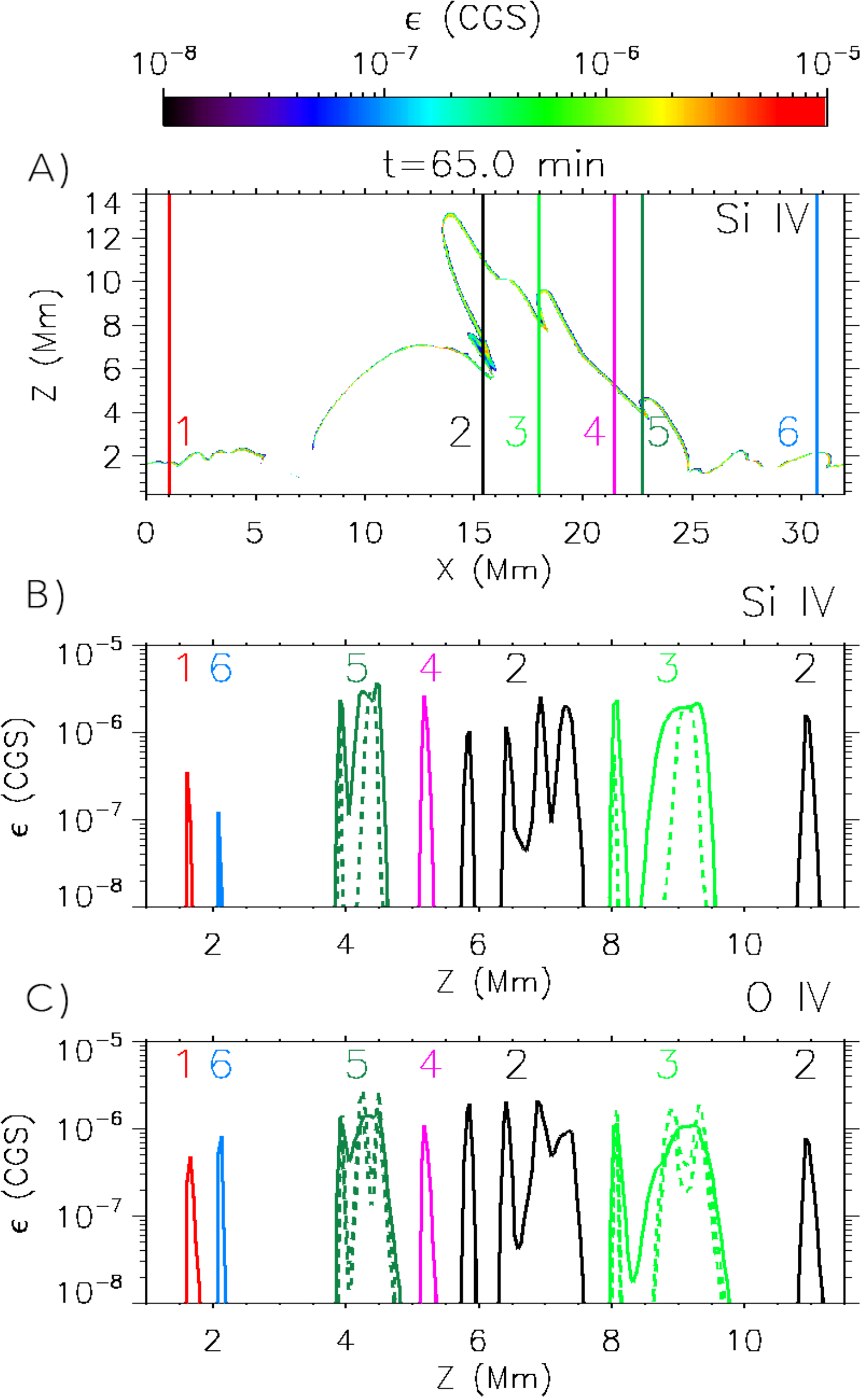}
\caption{Illustration of the multiple crossings of the transition
    region by a single LOS. A) 2D map of the \SiIV\ emissivity with colored
    and labelled lines overplotted in regions of interest. The lines
    corresponds to different vertical cuts used in panels B and C. B)
    \SiIV\ emissivity versus the height $Z$ for the different vertical cuts
    shown in the 2D map. C) Like panel B but for the \OIV\ 
    emissivity. Additionally, for the cuts labelled 3 and 5 we have added as a dashed line the
    corresponding emissivity if SE had been assumed.
} \label{figure11}
\end{figure}

To further illustrate those two effects, we use Figure
  \ref{figure11}.  The top panel contains a 2D map of the \SiIV\ emissivity
in which vertical cuts in different regions of interest are overplotted
through colored and labelled lines.  The corresponding \SiIV\ emissivity
distribution along those cuts is shown in panel B. Additionally, a similar
plot but for the \OIV\ distribution is shown in panel C. Those vertical
cuts are also shown in panels B and C of Figure \ref{figure10} for comparison
purposes. The light and dark
green cuts (numbers 3 and 5) are typical examples of the effect of
the tangency between the LOS and the surge's TR at the crests: note the
enhanced width of the maximum on the right in those two cuts due to tangency
effects. Those cuts and
the one in black (number 2) further show the effect of
multiple crossings of the TR by the LOS. In particular, the LOS
drawn in black crosses the folded TR near the internal footpoint four times
($5.5 < z < 8$ Mm), and an additional time at the the top of the surge ($z
\sim 11$ Mm). For the sake of completeness, we have added a dashed
line for the crests (3 and 5) in the middle and lower panels showing the
emissivity if SE had been assumed: the thickness of the high-emissivity TR
would be much smaller.  Finally, in contrast to all
the foregoing cases, the rest of the surge (pink line, label 4) and the QTR
(red and blue lines, labels 1 and 6, respectively) show a simpler geometry,
there is no TR-LOS alignment and there is just a single crossing.

%
%
\Needspace{5\baselineskip}
\section{Discussion}\label{sec:6}
In this paper, we have carried out two 2.5D radiative-MHD numerical experiments 
of magnetic flux emergence through granular convection and into the solar
atmosphere. The experiments were performed with Bifrost, including an 
extra module of the code that computes the nonequilibrium ionization (NEQ)
of silicon and oxygen. The time evolution of the two experiments leads to the 
formation of a cool and dense surge. We have studied the relevance
of the NEQ ionization for the presence of \SiIV\ and \OIV\ in the
periphery of the surge 
and how it affects the corresponding emissivities. The properties
 of the surge plasma emitting in \SiIV\ and \OIV\ were then
characterized and compared 
with those of the general TR plasma outside of the surge. We
have also analyzed the role of the heat  
conduction and optically-thin radiative cooling in the NEQ ionization. 
Furthermore, through forward modelling, 
we have understood different features
of the synthetic spectral profiles of \SiIV\ and \OIV, explaining the importance of the
shape of the transition region surrounding the surge
combined with the different possible angles of the LOS and providing
predictions for future observational studies.

In the following, we first address the implications of the
importance of the NEQ ionization in 
numerical experiments of eruptive phenomena in which heating and 
cooling are key mechanisms. (Section \ref{sec:6.1}). We then discuss 
their relevance for present and future observations (Section
\ref{sec:6.2}).

%
%
\Needspace{5\baselineskip}
\subsection{On the importance of the nonequilibrium (NEQ) ionization}\label{sec:6.1}

The main result of this paper is that the envelope of the emerged
domain, more specifically, 
of the dome and surge, are strongly affected by NEQ ionization (Section \ref{sec:4.1}). 
Focusing on the boundaries of the surge, comparing the number densities of emitters
computed via detailed solution of the NEQ rate equations with those obtained 
assuming statistical equilibrium (SE) we have concluded that the SE assumption 
would produce an erroneous result in the population of \SiIV\ and \OIV, mainly 
because it leads to a gross underestimate of the number density of emitters. The transition 
region outside of the flux emergence site is also affected by NEQ, but to a smaller 
extent. 

The above result has consequences in the corresponding emissivity (Section
\ref{sec:4.2}) and therefore, in the interpretation of the observations.  By
means of statistical analysis, we have shown that the boundaries of the surge
have greater values of the \SiIV\ emissivity than the region outside of the
flux emergence site.  Correspondingly, we have given the name
\textit{enhanced transition region} (ETR) to the former and \textit{quiet
  transition region} (QTR) to the latter (Section \ref{sec:4.2.1}).  This
difference is part of the explanation of why the surge is a brighter
structure than the rest of the transition region in the \IRIS\ observations
by NS2017. Furthermore, the joint probability distributions for emissivity
and temperature are not centered at the peak formation temperature of
\SiIV\ or \OIV\ in SE (see $T_{_{SE}}$ in Table \ref{table1} and Section
\ref{sec:4.2.2}). This reinforces earlier results
  (e.g., \citealp{olluri:2015}) about the inaccuracies inherent in the process of deducing
  temperature values from observations in transition region lines using SE
  considerations.

In Section \ref{sec:4.3} we have found that there are two different
populations concerning the thermal evolution that leads to the \SiIV\ and
\OIV\ emissivity. They have very different origins (one in the emerged plasma
dome, the other in coronal heights) but both are characterized by the
fact that they go through a period of rapid thermal change caused by the
optically thin losses and thermal conduction: the maximum values of the
\SiIV\ and \OIV\ emissivity in them are related to short characteristic
cooling times: $20-100$ s for $\tau_{thin}$ and around $4-40$ s for
$\tau_{Spitz}$. Those characteristic times are compatible with the
theoretical results by \cite{Smith:2010}, who found that for typical
densities of the active corona and the transition region, the solar plasma
can be affected by NEQ effects if changes occur on timescales shorter than
about $10-100$ s.  Those results highlight the role of optically thin losses
and thermal conduction because a) they provide the physical mechanism to
diminish the entropy and, consequently, obtain plasma with the adequate
temperatures to form ions of \SiIV\ or \OIV\ ($\sim 10^5$ K ); and b) they
are fast enough to produce important departures from SE. On the other hand,
the ion populations calculated through the present NEQ module in Bifrost are
not used in the energy equation of the general R-MHD calculation (see Section
\ref{sec:2.1}), so this could underestimate the effects of the entropy sinks
in the experiments. In fact, \cite{Hansteen:1993} found deviations of more
than a factor two in the optically thin losses when considering
nonequilibrium effects in his loop model, so $\tau_{thin}$ could be even more
efficient.

Our results indicate that surges, although traditionally described as
chromospheric phenomena, show important emission in transition region lines
due to the NEQ ionization linked to the quick action of the cooling
  processes, so the response of the transition region is intimately tied to
the surge dynamics and energetics. In fact, the same statement may apply for
other eruptive phenomena, in which impulsive plasma heating and cooling
occurs \citep[see][for a review of NEQ processes in the solar
  atmosphere]{Dudik:2017rv}.

%
%
\Needspace{5\baselineskip}
\subsection{Understanding observations and predictions for the future}\label{sec:6.2}

  From the number density of emitters and emissivity
  results of Section~\ref{sec:4} we gather that calculating heavy element
  populations directly through the rate equations instead of via the
  assumption of statistical equilibrium can be important to understand the
  observations of surges (and of other fast phenomena which reach TR
  temperatures in their periphery like spicules \citealt{DePontieu:2017l1},
  or UV bursts \citealt{Hansteen:2017ib}). In that section, analyzing the \SiIV\
  and \OIV\ emissivities of the plasma elements, we find that the ratio between them is approximately 2 in the regions with the highest emission within the ETR. Even though the intensity ratio is more commonly used (\citealp{Hayes:1987,Feldman:2008,Polito2016}, among others), the emissivity ratio can also be a valuable tool to understand the behavior of the ions in different regions of the Sun. In particular, we see that in the ETR this ratio is larger than in the transition region that has not been perturbed by the
flux emergence and subsequent surge and/or jet phenomena (QTR).
  
  To provide theoretical support to the NS2017
observations and predictions for future ones, in Section \ref{sec:5} we have
  therefore computed the synthetic profiles of \SiIV\ 1402.77
\AA\ and \OIV\ 1401.16 \AA, taking into account the Doppler shift because of
the plasma velocity and degrading the spatial and spectral resolutions to the
\IRIS\ ones.  A line-of-sight
  integration of the emissivities has also been carried out, to provide a
  measure for the total intensity emitted by the different regions of the
  surge. The strongest brightenings in \SiIV\ and \OIV\ have been located at
  the site of the internal footpoint, followed by the crests and the external
  footpoint (Figure \ref{figure9}, Section \ref{sec:5.1}). The intensity
  ratio between \SiIV\ and \OIV\ in those regions is, approximately, 5 (although it
  can range from 2 and 7).  Those values are between those for a coronal hole 
  and a quiescent active region obtained by \cite{Martinez-Sykora:2016obs}, which
  is consistent since we are mimicking an initial stratification similar to a coronal
  hole in which a total axial magnetic flux in the range of an 
  ephemeral active region \citep{Zwaan:1987yf} has been injected. The comparison
  of the total intensity for the NEQ and SE cases (Figure \ref{figure10},
  Section \ref{sec:5.1}) leads to a further indication of the importance of
using NEQ equations to determine the number density of \SiIV: the NEQ
calculation yields intensities coming out of the surge which are a factor
between $2$ and $10$ larger than when SE is assumed. For \OIV, instead, the
NEQ and SE calculations yield similar integrated intensities.  For \OIV, therefore, 
the NEQ calculations are important mainly to determine derived quantities like 
number densities of emitters $n_u$ (Section \ref{sec:4.1}) and temperatures (Section \ref{sec:4.2}).
In addition, we have found that for \SiIV\ all the regions in the surge have
a greater intensity than the QTR: this can explain why the surge
can be observationally distinguished from the QTR, as found in
the NS2017 paper. 

The high brightness of various features in the surge has been seen
  to arise in no small part from different LOS effects tied to the peculiarly irregular shape
  of its TR, and, in particular, to its varying inclination and the folds
  that develop in it (Section~\ref{sec:5.2}). On the one hand, whenever LOS
  and tangent plane to the TR are not mutually orthogonal, the issuing
  intensity collects emissivity from a larger number of plasma elements in
  the TR (alignment effect); on the other hand, given that the surge's TR is
  variously folded, forming crests and wedges, the LOS crosses the emitting
  layer multiple times (multiple-crossing effect). The alignment and
  multiple-crossing effects are quite evident in the footpoints and
  crests. This explains their remarkable brightness and makes clear their
  potential as beacons to indicate the presence of those special features in
  surge-like phenomena when observed in TR lines like \SiIV\ (Figure
  \ref{figure11}). Additionally, the multiple crossings can also have an
impact on the observed Doppler shifts since we could be integrating various
emitting layers with different dynamics. So, when confronted with a TR 
observation of a region where a surge is
  taking place, the detection of strong brightenings in
\SiIV\ and \OIV\ could help unravel the geometry of
the surge. Furthermore, since the internal footpoint of
the surge is close to the reconnection site, we might also find observational
evidences of reconnection in the neighborhood.  This provides theoretical
support to the location of the brightenings in the \IRIS\ observations by
NS2017. In addition, if strong signals are observed in \SiIV\ related to some
moderate signal in \OIV, they could correspond to the crests and the external
footpoint of the surge; nonetheless, the rest of the surge structure could be
only differentiated from the transition region in \SiIV.

%
%
\ \vspace{-2mm} 
\acknowledgments We gratefully acknowledge financial support
by the Spanish Ministry of Economy and Competitiveness (MINECO) through
projects AYA2011-24808 and AYA2014-55078-P, as well as by NASA through 
grants NNX16AG90G, NNH15ZDA001N, NNX17AD33G, and by NSF grant 
AST1714955 and contract NNG09FA40C (\IRIS).
We also acknowledge the computer resources and assistance provided at the 
MareNostrum (BSC/CNS/RES, Spain) and TeideHPC (ITER, Spain) 
supercomputers, where the calculations have been carried out, and at the
Pleiades cluster (computing projects s1061, s1472 and s1630 from the High 
End Computing division of NASA), where relevant code developments
have been made. Finally, the authors are grateful to Dr. Peter R. Young for his 
suggestions during the \textit{Hinode}-11/\IRIS-8 science meeting,
and also to Dr.~Jaroslav Dud\'ik for his constructive comments to improve the paper.

%

\bibliographystyle{apj} \bibliography{collectionbib}

\end{document}